\documentclass{gGAF2e}

\usepackage[T1]{fontenc}
\graphicspath{{./fig/}{./png/}}
\usepackage{color}

\newcommand{\alphaem}{\ensuremath{\alpha_{\rm em}}}

\newcommand{\MeV}{\,{\rm MeV}}

\def\kB{k_{\rm B}}

\newcommand{\nab}{\bm{\nabla}}
\newcommand{\AAA}{\bm{A}}
\newcommand{\BB}{\bm{B}}

\newcommand{\EE}{\bm{E}}
\newcommand{\JJ}{\bm{J}}
\newcommand{\UU}{\bm{U}}

\newcommand{\ff}{\bm{f}}

\newcommand{\SSSS}{\mbox{\boldmath ${\sf S}$} {}}
\newcommand{\meanrho}{\overline{\rho}}

{}
{}
{}

{}
{}
{}
{}
{}
{}
{}
{}
{}

{}
{}
{}
{}
{}
{}
{}
{}
{}

{}
{}
{}

\newcommand{\meanv}{\overline{v}}
\newcommand{\EQA}{\begin{eqnarray}}
\newcommand{\ENA}{\end{eqnarray}}

\def\oA{\omega_{\rm A}}

\newcommand{\pencilc}{{\sc Pencil Code}}

\newcommand{\mnras}{Mon.\ Not.\ Roy.\ Astron.\ Soc.}
\newcommand{\prd}{Phys.\ Rev. D}

\newcommand{\apj}{Astrophys.\ J.}

\newcommand{\aap}{Astron.\ \& Astrophys.}

\renewcommand{\max}{{\rm max}}

\def\half{{\textstyle{1\over2}}}
\def\onethird{{\textstyle{1\over3}}}

\def\Pm{{\rm Pr}_{_\mathrm{M}}}
\def\Rm{{\rm Re}_{_\mathrm{M}}}

\newcommand{\PrM}{{\rm Pr}_{_{\rm M}}}

\usepackage{verbatim,url}

\begin{document}

\jvol{00} \jnum{00} \jyear{2012} 

\markboth{J.~Schober et al.}{GEOPHYSICAL AND ASTROPHYSICAL FLUID DYNAMICS}

\articletype{GAFD Special issue on ``Physics and Algorithms of the Pencil Code''}

\title{{\textit{Chiral fermion asymmetry in high-energy plasma simulations}}}

\author{
J. SCHOBER${^1}$$^{\ast}$\thanks{$^\ast$Corresponding author. Email: jennifer.schober@epfl.ch},
A. BRANDENBURG${^{2,3,4}}$, \&
I. ROGACHEVSKII${^{5,2}}$\\
\vspace{6pt}  ${^1}$ Laboratoire d'Astrophysique, EPFL, CH-1290 Sauverny, Switzerland\\
\vspace{6pt}  ${^2}$ Nordita, KTH Royal Institute of Technology
 and Stockholm University, Roslagstullsbacken 23,
 10691 Stockholm, Sweden \\
\vspace{6pt}  $^3$JILA and Laboratory for Atmospheric and Space Physics,
    University of Colorado, Boulder, CO 80303, USA\\
\vspace{6pt}  $^4$Department of Astronomy, AlbaNova University Center,
    Stockholm University, SE-10691 Stockholm, Sweden\\
\vspace{6pt}  $^5$Department of Mechanical Engineering,
 Ben-Gurion University of the Negev, P.O. Box 653, Beer-Sheva
 84105, Israel
}

\received{\today,~ $ $Revision: 1.245 $ $}

\maketitle

\begin{abstract}
The chiral magnetic effect (CME) is
a quantum relativistic effect that describes the appearance
of an additional electric current along a magnetic field.
It is caused by an asymmetry between the number densities
of left- and right-handed fermions, which
can be maintained at high energies when the chirality
flipping rate can be neglected,
for example in the early Universe.
The inclusion of the CME in the Maxwell equations leads to a modified set of
magnetohydrodynamical (MHD) equations. The CME is studied here in numerical
simulations with the \pencilc.
We discuss how the CME is implemented in the code and how the time step and
the spatial resolution of a simulation need
to be adjusted in presence of a chiral asymmetry.
The CME plays a key role in the evolution of magnetic fields, since it
results in a dynamo effect associated with an additional term
in the induction equation.
This term is formally similar to the
$\alpha$ effect in classical mean-field MHD.
However, the chiral dynamo can operate without turbulence and is
associated with small spatial scales that can be, in the case of the early
Universe, orders of magnitude below the Hubble radius.
A chiral $\alpha_\mu$ effect has also been identified
in mean-field theory.
It occurs in the presence of turbulence, but is not related to kinetic helicity.
Depending on the plasma parameters, chiral dynamo instabilities can
amplify magnetic fields over many orders of magnitude.
These instabilities can potentially affect the propagation of MHD waves.
Our numerical simulations demonstrate strong modifications of the dispersion
relation for MHD waves for large chiral asymmetry.
We also study the coupling between the
evolution of the chiral chemical potential and the ordinary
chemical potential, which is proportional to the sum of the number densities of left- and
right-handed fermions.
An important consequence of this coupling is the emergence of chiral magnetic waves (CMWs).
We confirm numerically that linear CMWs and MHD waves are not interacting.
Our simulations suggest that the chemical potential has only a
minor effect on the non-linear evolution of the chiral dynamo.


\end{abstract}

\section{Introduction}
\label{sec_intro}

Research in turbulence physics was always strongly guided by input
from experiments and also astronomical observations.
This also applies to magnetohydrodynamic (MHD) turbulence,
studied in solar and space physics, astrophysics, as well as in
liquid sodium experiments \citep{Gal00,SM01,Monchaux07}.
These investigations corroborate the existence of the $\alpha$ effect,
which enables a large-scale dynamo caused by helical turbulent motions
\citep{M78,KR80,ZRS83}.
In recent times, MHD turbulence simulations have played important roles
in demonstrating various scaling laws that cannot easily be determined
observationally.
However, under the extreme conditions of the early universe or in neutron
stars, for example, only very limited information about the nature of such
turbulence is available.
Here, numerical simulations play a particularly crucial role.
They allow new physical effects to be modeled and studied under
turbulent conditions.

The \pencilc\footnote{\url{https://github.com/pencil-code}, DOI:10.5281/zenodo.2315093} is designed for
exploring the dynamical evolution of turbulent,
compressible, and magnetized plasmas in the MHD limit.
It is, in particular, suitable for studying a large variety of
cosmic plasmas and astrophysical systems
from planets and stars, to the interstellar medium, galaxies, the
intergalactic medium, and cosmology.
In its basic configuration, the
\pencilc\ solves the equations of classical MHD, which describe the evolution
of the mass density, $\rho$, the magnetic field strength, $\BB$, the velocity,
$\UU$, and the temperature, $T$.
Interestingly, this set of dynamical variables has to be extended in the limit
of high energies, where a new degree of freedom, the
\textit{chiral chemical potential}, arises from the chiral magnetic
effect (CME).
This anomalous fermionic quantum effect emerges within the standard model
of high energy particle physics and describes the generation of
an electric current along the magnetic field
if there is an asymmetry between the number density of
left- and right-handed fermions.
The CME modifies the Maxwell equations
and leads to a system of chiral MHD equations,
which turn into classical MHD when the chiral chemical potential vanishes.
In this paper, we describe how the CME affects
a relativistic plasma and how it can be explored with a new module in
the \pencilc.

The CME was first
suggested by \citet{Vilenkin:80a} and was later derived independently by
\citet{NielsenNinomiya83}.
These findings triggered many
theoretical studies of the effect in various fields,
from cosmology \citep{Joyce:97,Semikoz:04a,TVV12,BFR12,Boyarsky:15a,DvornikovSemikoz2017}
and neutron stars \citep{DvornikovSemikoz2015,SiglLeite2016,Yamamoto:2016xtu},
to heavy ion collisions \citep{Kharzeev:2013ffa,Kharzeev:2015znc} and
condensed matter \citep{Miransky:2015ava}.
Some of the theoretical predictions have already been confirmed experimentally
in condensed matter \citep{Wang13,ALICE13}.
Three dimensional high-resolution direct numerical simulations (DNS) are an
additional tool for gaining deeper understanding of the importance of the
CME in high energy plasmas.
Therefore, a new module for chiral MHD has been implemented in the \pencilc.
The module is based on a system of equations that has been
derived by \citet{REtAl17}.
An important extension of those equations is,
however, the inclusion of the evolution of the
ordinary (achiral) chemical potential,
which is proportional to the sum of the number densities of
left- and right-handed fermions.
Previous investigations have demonstrated that a non-vanishing chiral chemical potential can result
in chiral MHD dynamos, which have later been confirmed in DNS \citep{SRBBFRK17}.
One important implication of chiral MHD dynamos is the generation of
chiral-magnetically driven turbulence with an energy spectrum proportional
to $k^{-2}$ within well-defined boundaries in wavenumber $k$
\citep{BSRKBFRK17,SBRK18}.

In this paper we discuss the implementation of chiral MHD in the \pencilc\
which is, as far as we know, one of the first codes that includes a full implementation of the
CME in the MHD limit; but see also \citet{MKTY18} and \citet{DZB18} for more recent examples of other codes.
In section~\ref{sec_chiralMHD}, we provide an introduction to the physical
background of the CME and highlight the most important properties of the
set of chiral MHD equations in terms of numerical modelling.
The implementation of chiral MHD in the \pencilc\ is described in
section~\ref{sec_implementation}.
In section~\ref{sec_usage} we discuss how chiral MHD can be explored
in DNS and what to expect in different exemplary numerical scenarios.
We discuss chiral MHD dynamos, effects of turbulence, the modification
of MHD waves, and finally chiral magnetic waves caused by a non-zero chemical
potential.
We draw our conclusions in section~\ref{sec_conclusions}.

\section{Theoretical background}
\label{sec_chiralMHD}

\subsection{The nature of the CME}

\begin{figure}
\begin{center}
  \includegraphics[width=0.3\textwidth]{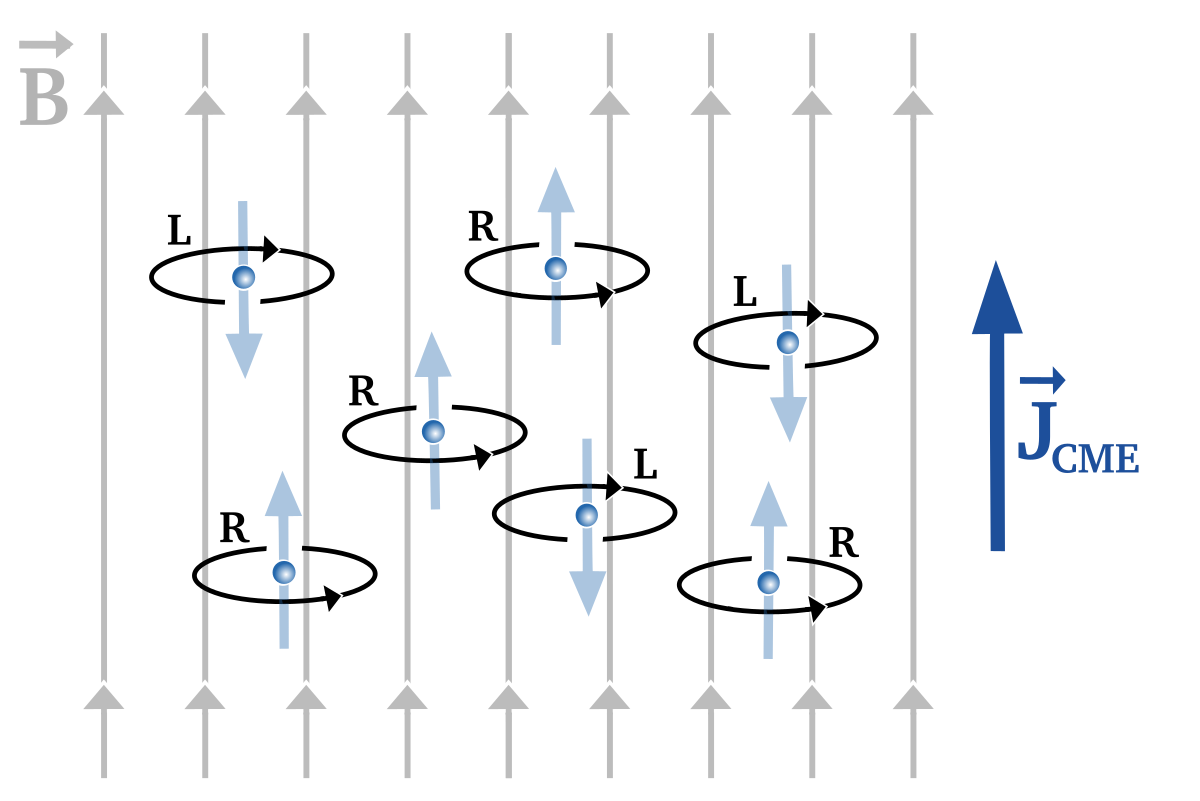}
\caption{Illustration of the chiral magnetic effect. In the presence of an
external magnetic field $\BB$, the momenta of right-(left-)handed
fermions at the lowest Landau level align with
(antiparallel to) the field lines.
An asymmetry between the number densities of left- and right-handed particles
results in a net electric current
$\JJ_{\rm CME}$. (Colour online)}%
\label{fig_CME}
\end{center}
\end{figure}

The CME occurs in magnetized relativistic plasmas, in
which the number density of left-handed fermions
differs from the one of right-handed
fermions \citep[see e.g.][for reviews]{KLSY13,Kharzeev:2013ffa,Kharzeev:2015znc}.
This asymmetry is described by the chiral chemical potential\footnote{
The notation with the number $5$ indicates that $\mu_5$ arises from quantum mechanics.
Here, a Dirac field can be projected onto its left- and right-handed components
using $\gamma^5 \equiv i \gamma^0 \gamma^1 \gamma^2 \gamma^3$, where $\gamma^n$
with $n=0,1,2,3$ are the Dirac matrices.}
\begin{eqnarray}
  \mu_5^\mathrm{phys} \equiv \mu_{_{\rm L}}^\mathrm{phys} -\mu_{_{\rm R}}^\mathrm{phys},
\label{eq_mu5phys}
\end{eqnarray}
which is defined as the difference between the chemical potential of
left- and right-handed fermions, $\mu_{_{\rm L}}^\mathrm{phys}$ and
$\mu_{_{\rm R}}^\mathrm{phys}$, respectively.\footnote{
The superscript ``phys'' indicates that the chemical potential is
given in its usual physical dimension of energy;
the symbol $\mu_5$ will later be used for a rescaled chiral chemical potential.}
In the presence of a magnetic field, the momentum vectors
of the fermions at the lowest Landau level align with
the field lines while their direction depends on the handedness of the fermion;
see the illustration in figure~\ref{fig_CME}.
A non-vanishing $\mu_5^\mathrm{phys}$ leads to the occurrence of
the electric current
\begin{eqnarray}
    \JJ_{\rm CME} = \frac{\alphaem}{\pi \hbar} \mu_5^\mathrm{phys} \BB ,
  \label{eq_CME}
\end{eqnarray}
where $\alphaem \approx 1/137$ is the fine structure constant
and $\hbar$ is the reduced Planck constant
\citep{Vilenkin:80a,Alekseev:98a,Frohlich:2000en,Fukushima:08,Son:2009tf}.
The presence of $\alphaem$ indicates that the CME is a quantum effect.

\subsection{System of chiral MHD equations}

The chiral electric current~(\ref{eq_CME})
adds to the classical Ohmic current, leading
to a modification of the Maxwell equations.
Combining these equations with Ohm's law,
the following set of chiral MHD equations
for an isothermal high temperature plasma is obtained:
\begin{eqnarray}
  \frac{\upartial \BB}{\upartial t} &=& \nab   \times   \left[{\UU}  \times   {\BB}
  - \eta \, \left(\nab   \times   {\BB}
  - \mu_5 {\BB} \right) \right] ,
\label{ind-DNS}\\
  \rho{{\mathrm D} \UU \over {\mathrm D} t}&=& (\nab   \times   {\BB})  \times   \BB
  -\nab  p + \nab  {\bm \cdot} (2\nu \rho \SSSS)
  +\rho \ff ,
\label{UU-DNS}\\
  \frac{{\mathrm D} \rho}{{\mathrm D} t} &=& - \rho \, \nab  \cdot \UU ,
\label{rho-DNS}\\
  \frac{{\mathrm D} \mu_5}{{\mathrm D} t} &=& D_5 \, \Delta \mu_5
  + \lambda \, \eta \, \left[{\BB} {\bm \cdot} (\nab   \times   {\BB})
  - \mu_5 {\BB}^2\right]
  - C_5 ({\BB} {\bm \cdot} \nab) \mu
  -\Gamma_{\rm\!f}\mu_5,
\label{mu5-DNS} \\
  \frac{{\mathrm D} \mu}{{\mathrm D} t}  &=& D_\mu \, \Delta \mu - C_\mu (\BB {\bm \cdot} \nab)  \mu_5 ,
\label{mu-DNS}
\end{eqnarray}
where the magnetic field $\BB$ is normalised such that the magnetic energy
density is $\BB^2/2$ (so the magnetic field in Gauss is $\sqrt{4\pi}\,\BB$),
$\eta$ is the magnetic resistivity, $\UU$ is the velocity, and
${\mathrm D}/{\mathrm D} t \equiv \upartial/\upartial t + \UU \cdot \nab$.
The normalised chiral chemical potential
$\mu_5 \equiv (4 \alphaem /\hbar c) \mu_5^\mathrm{phys}$ is scaled such that
it has the same units as a wavelength (inverse length); see also table~\ref{tab_chiralMHDunits}.
The chiral nonlinearity parameter is
\begin{eqnarray}
  \lambda=3 \hbar c \left({8 \alphaem \over \kB T} \right)^2,
\label{eq_lambda}
\end{eqnarray}
where $T$ is the temperature, $\kB$ is the Boltzmann constant and
$c$ is the speed of light.
The expression for $\lambda$ given above is valid for
$\kB T \gg \max(|\mu_{_{\rm L}}^\mathrm{phys}|,|\mu_{_{\rm R}}^\mathrm{phys}|)$, which holds for the
description of the hot plasma in the early Universe.
In a dense plasma, like within a neutron star where
$\kB T \ll \max(|\mu_{_{\rm L}}^\mathrm{phys}|,|\mu_{_{\rm R}}^\mathrm{phys}|)$, a dependence on the usual
chemical potential $\mu^\mathrm{phys} \equiv\mu_{_{\rm R}}^\mathrm{phys} +\mu_{_{\rm L}}^\mathrm{phys}$
needs to be included
\citep[see, e.g.][]{Kharzeev:2013ffa,Kharzeev:2015znc,DvornikovSemikoz2015}.

In equations~(\ref{ind-DNS})--(\ref{mu-DNS}),
$p$ is the fluid pressure,
${\sf S}_{ij}=\half(U_{i,j}+U_{j,i})-\onethird\delta_{ij} {\bm \nabla}
{\bm \cdot} \UU$
are the components of the trace-free strain tensor, where commas denote partial
spatial differentiation, $\nu$ is the kinematic viscosity,
and $\ff$ is a forcing function used to drive turbulence in DNS.
For an isothermal equation of state, the pressure $p$ is related
to the mass density $\rho$ via $p=c_{\rm s}^2\rho$, where $c_{\rm s}$ is the
isothermal sound speed.
The last term in equation~(\ref{mu5-DNS}) describes
the chiral flipping reactions at a rate $\Gamma_{\rm\!f}>0$.
This rate characterises the flipping between left- and right-handed states of
a fermion and becomes important at low temperatures, i.e.~when the mass
of the particles cannot be neglected anymore.
Equation~(\ref{mu-DNS}) describes the evolution of
$\mu^\mathrm{phys}$, which
is, in consistency with $\mu_5$, normalised as\footnote{
We note that in previous works \citep{REtAl17,BSRKBFRK17,SRBBFRK17},
the symbol ``$\mu$'' was used for the
normalised chiral chemical potential.
Due to the inclusion of the evolution of the ordinary chemical potential,
a change of notation became necessary for the extended chiral MHD
equations used in the present study.}
$\mu \equiv (4 \alphaem /\hbar c) \mu^\mathrm{phys}$.
Equation~(\ref{mu-DNS}) for $\mu$ and equation~(\ref{mu5-DNS})
for $\mu_5$ have been derived by \cite{G16} using chiral kinetic theory
in the high-temperature limit.
The evolution equations of $\mu_5$ and $\mu$
are coupled through the coupling
parameters $C_5$ and $C_\mu$, respectively, and
$D_5$ and $D_\mu$ are diffusion coefficients.

\begin{table}[b!]
  \centering
  \caption{Physical units in chiral MHD}
  \label{tab_chiralMHDunits}
  \begin{tabular}{llll}
  \hline
  \hline
     Parameter		 	& cgs unit 	     &  Natural unit   & Comment    \\
  \hline
  	$\mu_5^\mathrm{phys}$   & erg							& eV			&	 \\
  	$\mu_5$      		& $\mathrm{cm}^{-1}$					& eV			&  $\mu_5 = 4 \alphaem/(\hbar c) \mu_5^\mathrm{phys}$  \\
  	$\mu^\mathrm{phys}$   	& erg							& eV			&	 \\
  	$\mu$      		& $\mathrm{cm}^{-1}$					& eV			&  $\mu = 4 \alphaem/(\hbar c) \mu^\mathrm{phys}$  \\
	$\lambda$		& $\mathrm{s}^{2}\mathrm{g}^{-1}\mathrm{cm}^{-1}  $ 	& $\mathrm{eV}^{-2}  $	& 	\\
        $D_5$			& $\mathrm{cm}^{2}\mathrm{s}^{-1}$			& $\mathrm{eV}^{-1} $	& 	\\
        $D_\mu$			& $\mathrm{cm}^{2}\mathrm{s}^{-1}$			& $\mathrm{eV}^{-1} $	& 	\\
        $\sqrt{4\pi}\BB$	& G=$\mathrm{g}^{1/2}\mathrm{cm}^{-1/2}\mathrm{s}^{-1}$	&  $\mathrm{eV}^{2}  $	&  defined such that $\BB^2/2$ is an energy density	\\
        $\Gamma_\mathrm{f}$	& $\mathrm{s}^{-1}$					&  $\mathrm{eV}  $	&  	\\
  \hline
  \hline
  \end{tabular}
\end{table}

In the \pencilc, a dimensionless form of the system of
equations~(\ref{ind-DNS})--(\ref{mu-DNS}) has been implemented.
We give this system of equations in Appendix~\ref{appendix_dimform}.
In the following we will use the chiral velocity, defined as
$v_\mu\equiv\eta\mu_{5,0}$, where $\mu_{5,0}\equiv\mu_5(t=0)$,
and the corresponding dimensionless chiral Mach number
${\rm Ma}_\mu\equiv v_\mu/c_\mathrm{s}$.
Since $\mu_5$ has the dimension of a wavenumber
(see also table~\ref{tab_chiralMHDparameters}), $v_\mu$ has the
dimension of a velocity.
Also, we introduce a dimensionless form of the chiral nonlinearity parameter
as $\lambda_5 = \lambda \eta^2 \meanrho$, where the overbar denotes a
volume average.
We note that the default setup of the \pencilc\ does not include the $\mu$
terms and equation~(\ref{mu-DNS}).
These terms can be switched on via the logical parameter \texttt{lmuS}.
If \texttt{lmuS=.true.}, the \texttt{MVAR CONTRIBUTION} in \texttt{cparam.local} needs to be increased
by one.

\subsection{Conservation law in chiral MHD}

A remarkable consequence of the system of equations~(\ref{ind-DNS})--(\ref{mu-DNS})
is that
\begin{equation}
\frac{\upartial }{\upartial t} \left({\lambda \over 2}  {\bm A} {\bm \cdot} \BB
+ \mu_5 \right) + \nab  {\bm \cdot}
\left[{\lambda \over 2} \left({\bm \EE} \times   {\bm A} - \BB \, \Phi\right)
-  D_5 \nab  \mu_5
+ C_5 {\BB} \mu
\right] = 0 ,
\label{CL}
\end{equation}
where $\EE = - \UU\times \BB + \eta \nab\times \BB - \eta \, \mu_5 \, \BB + {\rm O}(\eta^2)$
is the electric field and ${\bm A}$ is the magnetic vector potential,
with $\BB = \nab \times {\bm A}$;
see \citet{BFR12} and Section~4.3.\ of \citet{REtAl17}.
For periodic boundary conditions, which are often applied in MHD simulations,
the divergence term in equation~(\ref{CL}) vanishes
and hence $\lambda {\bm A} {\bm \cdot} \BB + 2 \mu_5 = \mathrm{const}$.
We stress that conservation of the sum of magnetic helicity density and chiral density
holds for arbitrary values of $\eta$.
This is different from classical MHD, where magnetic helicity
$\int {\bm A} {\bm \cdot} \BB \, dV$ is only conserved in the limit of
$\eta \to 0$.

From equation~(\ref{CL}), under the assumption
of vanishing initial magnetic helicity,
a maximum magnetic field strength for a
given initial chiral chemical potential
can be estimated through \citep{BSRKBFRK17}
\begin{equation}
  \overline{\BB^2}_\mathrm{sat}\,\xi_{\rm M} \approx {2 \mu_{5,0}}\big/{\lambda},
\label{eq_B2sat}
\end{equation}
where $\xi_{\rm M}$ is the correlation length of the magnetic field
and overlines denote volume averages.

\subsection{Length and time scales in chiral MHD}

\subsubsection{Laminar dynamo phase}

With a plane wave ansatz,
the linearised induction equation~(\ref{ind-DNS}) with the CME term and a
vanishing velocity field yields an instability that is characterized by
the growth rate
\begin{equation}
   \gamma(k) = |v_\mu k| - \eta k^2,
\label{eq_gammalam}
\end{equation}
with $k$ being the wavenumber.
The maximum growth rate of this instability is
\begin{eqnarray}
   \gamma_\mu = {v_\mu^2}\big/{(4 \eta)},
\label{eq_gammalam_max}
\end{eqnarray}
and the typical wavenumber of the dynamo instability in laminar flows is
\begin{equation}
   k_\mu = {|\mu_5|}\big/{2}.
\end{equation}
This chiral instability is caused by the term
$\nab\times(v_\mu \BB)$ in the
induction equation~(\ref{ind-DNS}) of chiral MHD.
We note that, while this term is formally similar to the $\alpha$ effect in
classical mean-field MHD, the $v_\mu$ is not produced by turbulence, but rather by a quantum
effect related to the handedness of fermions.
This is the $v_\mu^2$ dynamo.
In the presence of shear, its growth rate is modified in ways that are
similar to those of the classical
$\alpha\Omega$ dynamo \citep{REtAl17},
except that this chiral dynamo is not related to a turbulent flow.

\subsubsection{Turbulent dynamo phase}
\label{sec_turbulentflows}

In the presence of turbulence,
regardless of whether it is driven by a forcing function or by the Lorentz force,
the growth rate of the mean magnetic field obtained
in the framework of the mean-field approach \citep{REtAl17}, is given by
\begin{eqnarray}
   \gamma(k) = |(\meanv_\mu + \alpha_\mu)\, k| - (\eta+ \, \eta_{_{T}}) \, k^2,
\label{eq_gammaturb}
\end{eqnarray}
with $k^2=k_x^2 + k_z^2$ and $\meanv_\mu$ being the mean chiral chemical potential
multiplied by $\eta$.
In comparison to equation~(\ref{eq_gammalam}), turbulent diffusion
$\eta_{_{T}}= u_\mathrm{rms}/(3 k_\mathrm{f})$ adds to
Ohmic diffusion, where $k_\mathrm{f}$ is the forcing wavenumber.
Additionally, as has been shown by \citet{REtAl17}, the CME leads to a novel
large-scale dynamo that is caused by the $\alpha_\mu$ effect:
\begin{eqnarray}
  \alpha_\mu  =
  \begin{cases}
   - \dfrac{(q-1)}{3(q+1)} \, \Rm^2 \, \meanv_\mu ,& \mathrm{for}~\Rm\ll 1, \\[0.8em]
   -\dfrac{2}{3} \meanv_\mu \log\,{\Rm} ,& \mathrm{for}~\Rm\gg 1,
  \end{cases}
\label{eq_alpha}
\end{eqnarray}
with $1<q<3$.
$\Rm = u_\mathrm{rms}/(\eta k_\mathrm{f})$ is the magnetic Reynolds number.
The expression given in equation~(\ref{eq_alpha})
is valid for weak mean magnetic fields,
when the energy of the mean magnetic field is much smaller than the turbulent
kinetic energy.
While $\alpha_\mu$ is related to the fluctuations of the magnetic and velocity
field --
in contrast to the $\alpha$ effect in classical mean-field MHD -- kinetic helicity
is not required for it to occur.

The maximum growth rate of the mean magnetic field is
\begin{eqnarray}
   \gamma_\alpha = {(\meanv_\mu + \alpha_\mu)^2\over 4 (\eta+ \, \eta_{_{T}})}
                 = {(\meanv_\mu + \alpha_\mu)^2\over 4 \eta (1 + \, \Rm/3)}.
\label{eq_gammaturb_max}
\end{eqnarray}
The maximum growth rate of the $\alpha_\mu$ dynamo
is attained at the wavenumber
\begin{equation}
   k_\alpha = {|\meanv_\mu + \alpha_\mu|\over 2(\eta+ \, \eta_{_{T}})}
            = {|\meanv_\mu + \alpha_\mu|\over 2\eta\, (1 + \, \Rm/3)} .
\label{kmax_turb}
\end{equation}
provided that small-scale turbulence is present.

There is one more characteristic scale in chiral MHD turbulence, namely
the scale on which dynamo saturation occurs.
It has been shown in \citet{BSRKBFRK17} that, without applying a forcing
function in the Navier-Stokes equation, the CME produces
chiral-magnetically driven turbulence, which causes a $k^{-2}$ magnetic energy
spectrum between the wavenumbers $k_\mu$ and
\begin{equation}
  k_\lambda \approx 4 \sqrt{\meanrho\lambda}\,\mu_{5,0}\eta.
\label{klambda}
\end{equation}

Regarding spatial scales, we note that a fluid description, as presented here,
is only valid as long as all relevant chiral length scales are larger than the
mean free path.
Otherwise, a kinetic description of the plasma
\citep{Artsimovich-Sagdeev}
needs to be applied, which will not be discussed here.

\section{Application of the chiral MHD module in the {\bfseries{\scshape Pencil Code}} }
\label{sec_implementation}

\subsection{Implementation}

In comparison to classical MHD, in chiral MHD the evolution equation of at least one additional
scalar field, the chiral chemical potential $\mu_5(\boldsymbol{x},t)$, needs to be
solved\footnote{If $\mu$ is incorporated, two additional evolution equations
need to be solved.}.
The evolution equation for $\mu_5(\boldsymbol{x},t)$ is given by
equation~(\ref{mu5-DNS}).
Additionally, $\mu_5(\boldsymbol{x},t)$ enters the induction equation~(\ref{ind-DNS})
via the chiral dynamo term $\nab\times(\eta\mu_5\BB)$.

Chiral MHD is currently implemented in the \pencilc\ as a special module,
where $\mu_5(\mathbf{x},t)$ is made available as a pencil \texttt{p\%mu5}
and in the f-array, and
can be activated by adding the line
\begin{verbatim}
   SPECIAL   =   special/chiral_mhd
\end{verbatim}
to the file \texttt{src/Makefile.local}.
Obviously, also the \texttt{magnetic.f90} module needs to be switched on.
For solving the complete set of equations~(\ref{ind-DNS})--(\ref{mu-DNS}),
additionally, the \texttt{hydro.f90} module, the \texttt{density.f90} module,
and an equation of state module are required.
An example for the setup in the \pencilc\ is presented in the appendix.

\subsection{Time stepping}

The time step $\delta t$ in the \pencilc\ can either be set to a fixed value or be
adjusted automatically, depending on the instantaneous values
of characteristic time scales in the simulation.
In the latter case, $\delta t$ is specified by the Courant time step, which is taken
as the minimum of all involved terms of the equations solved in the simulation
and can be multiplied by a user-definable scale factor \texttt{cdt}
in the input file \texttt{run.in}.

The \texttt{chiral-mhd.f90} module introduces the following six time step
contributions:
\begin{align}
   \delta t_{\lambda_5} = \,&\frac{1}{\lambda \eta \BB^2}   ,\,&
   \delta t_{D_5} = \,&\frac{\delta x^2}{D_5}   ,\,&
   \delta t_{\Gamma_\mathrm{f}} =\,& \frac{1}{\Gamma_\mathrm{f}}, \label{eq_dt1} \\
\hskip15mm    \delta t_{\mathrm{CMW}} = \,&\frac{\delta x}{|\BB|\sqrt{C_5 C_\mu}},\,&
   \delta t_{D_\mu} =\,& \frac{\delta x^2}{D_\mu},\,&
   \delta t_{v_\mu} = \,&\frac{\delta x}{\eta \mu_5}.\hskip20mm \label{eq_dt2}
\end{align}
The contribution $\delta t_{\lambda_5}$ results from the term proportional to
$\lambda$ in equation~(\ref{mu5-DNS}) and $\delta t_{\Gamma_\mathrm{f}}$ from
the flipping term in the same equation.
The contributions $\delta t_{D_5}$ and $\delta t_{D_\mu}$ are required to
describe diffusion of $\mu_5$ and $\mu$, respectively.
Further, $\delta t_{v_\mu}$ results from the chiral dynamo term in the
induction equation and $\delta t_{\mathrm{CMW}}$ from chiral magnetic
waves (CMWs).
The total contribution to the time step calculated in the chiral MHD module,
is given by
\begin{eqnarray}
   \delta t_\mathrm{chiral} = c_\mathrm{\delta t,chiral} ~
     \mathrm{min} (\delta t_{\lambda_5}, \delta t_{D_5}, \delta t_{\Gamma_\mathrm{f}},
     \delta t_{\mathrm{CMW}}, \delta t_{D_\mu}, \delta t_{v_\mu}),
\end{eqnarray}
which can be scaled by the parameter $c_\mathrm{dt,chiral}$.
The default value of $c_\mathrm{dt,chiral}$ is chosen to be unity, but can
be set to smaller values in \texttt{run.in}.

\begin{figure}
\begin{center}
  \subfigure{\includegraphics[width=0.49\textwidth]{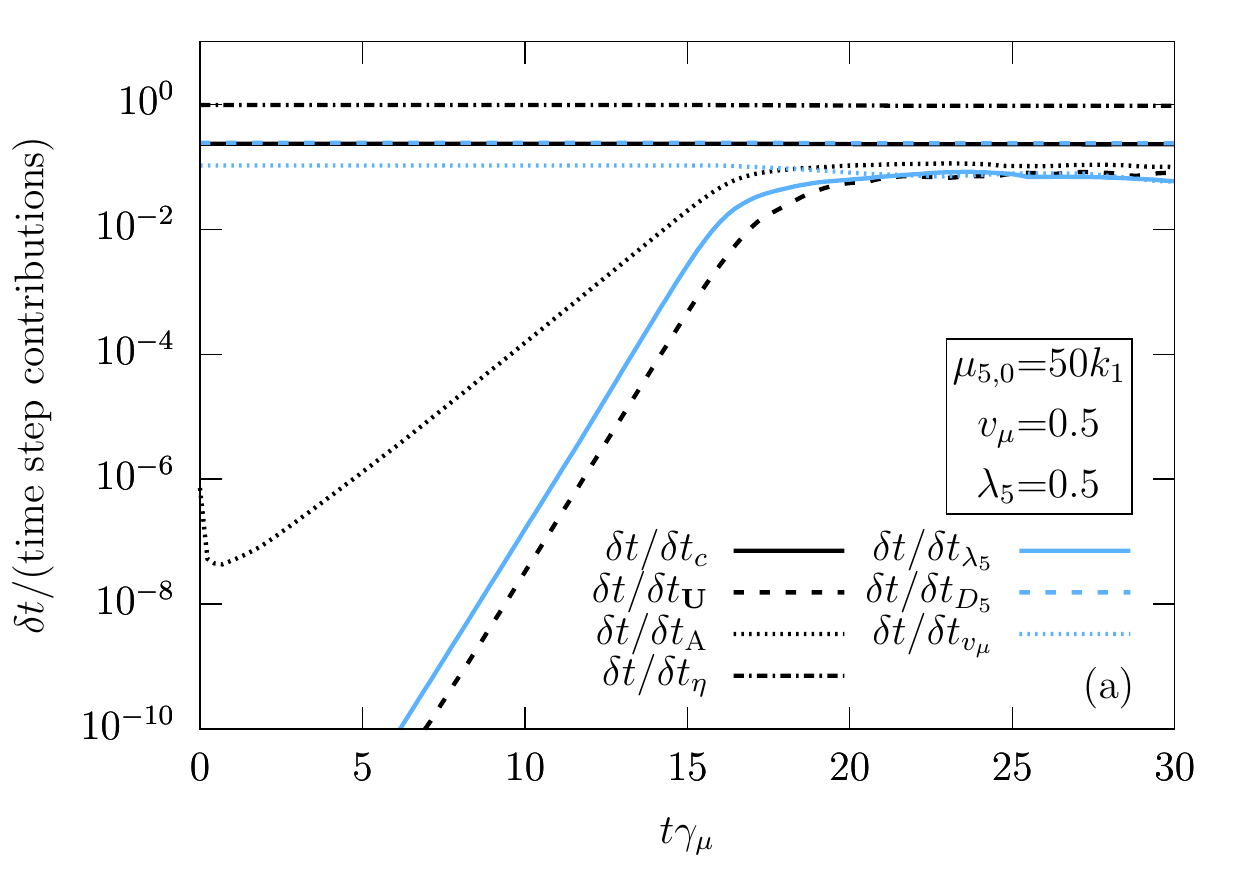}}
  \subfigure{\includegraphics[width=0.49\textwidth]{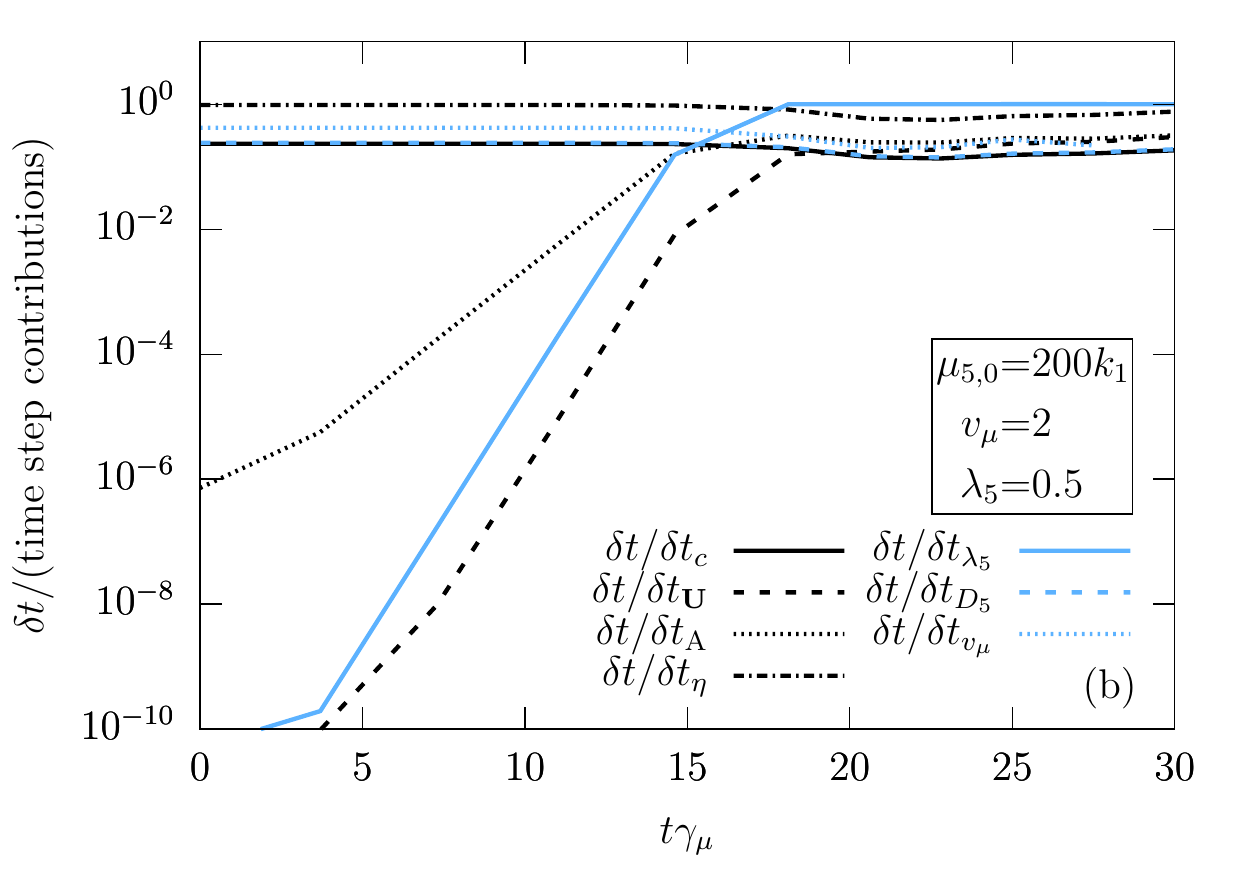}}
  \subfigure{\includegraphics[width=0.49\textwidth]{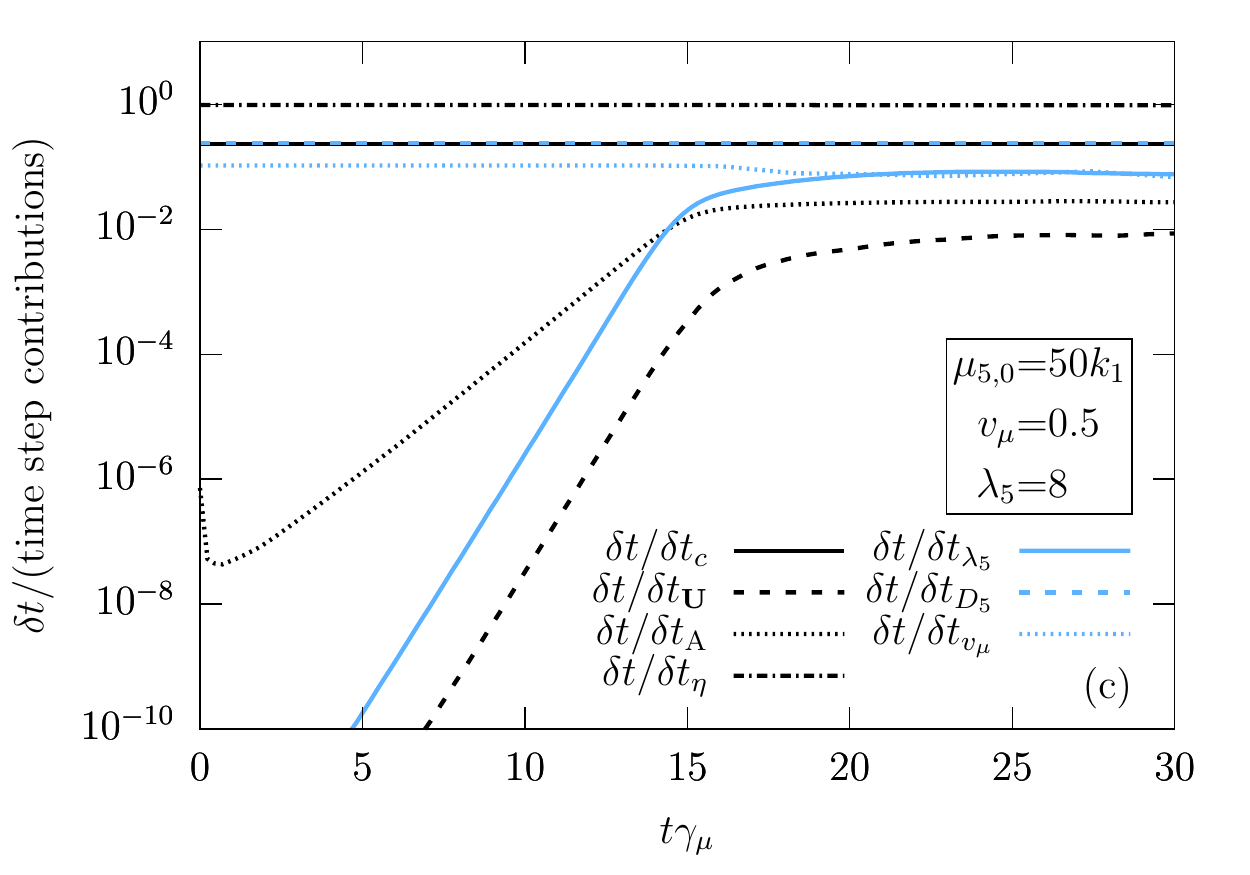}}
  \subfigure{\includegraphics[width=0.49\textwidth]{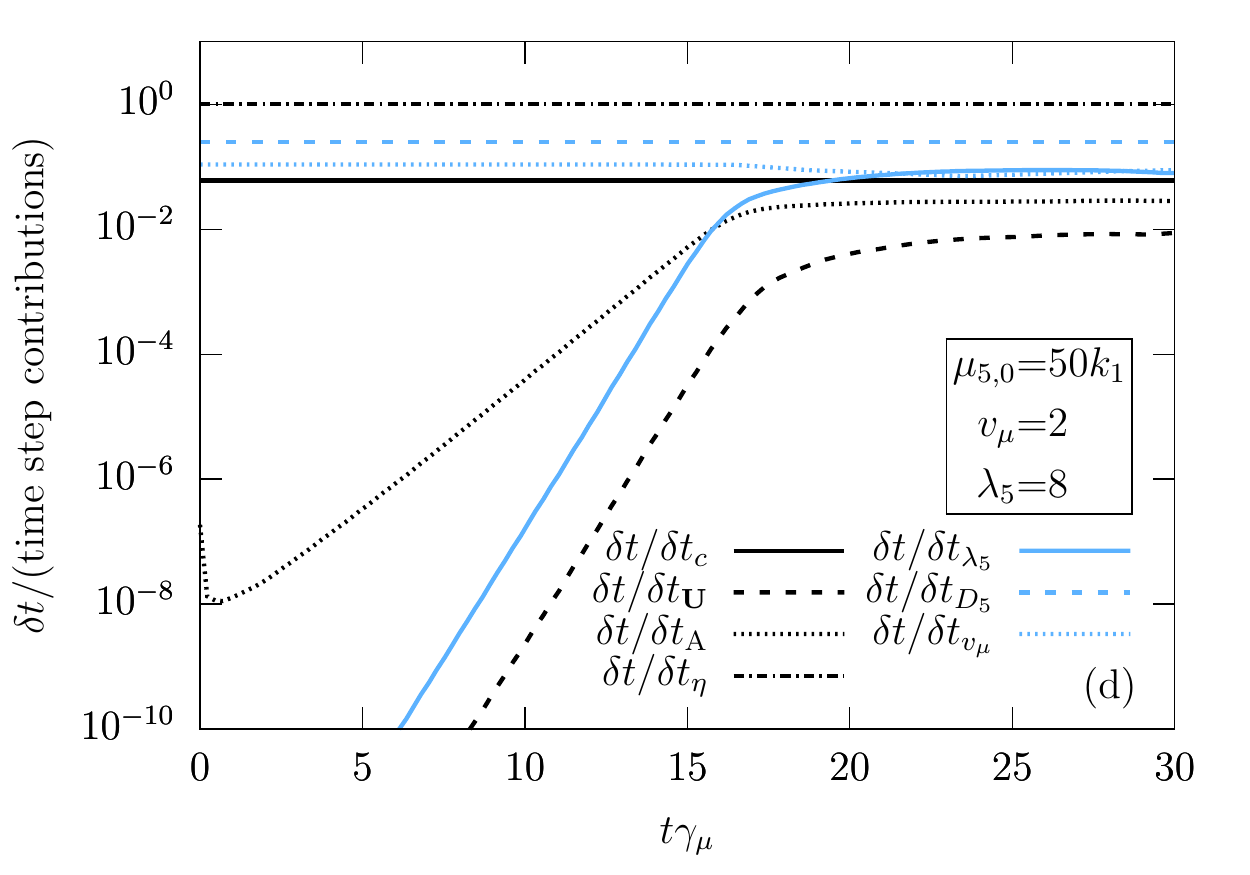}}
\caption{Comparison of various contributions to the time step.
Different panels show 2D simulations with different values of $\mu_{5,0}/k_1$, $v_\mu$, and
$\lambda_5$ as indicated in the boxes.
In all cases, $\Pm={\rm Pr}_{_{5}}=1$ and the resolution is $512^2$ mesh points.
Different lines show different time steps, where black colour indicates time
steps from classical MHD and blue colour shows time steps from the
chiral MHD module.
The time steps are normalised by the overall $\delta t$, which is determined as
the minimum of the individual time steps from all term involved. (Colour online)}
\label{fig_timestep}
\end{center}
\end{figure}

The relative importance of the chiral contributions to the simulation time
step is demonstrated in figure~\ref{fig_timestep}, where four different
simulations are presented.
These simulations are performed in two-dimensional (2D) domains with a size
of $(2\pi)^2$ and a resolution of $512^2$, that is, $\delta x\approx 0.012$,
and the magnetic and chiral Prandtl numbers are
$\Pm=\nu/\eta=1$ and ${\rm Pr}_{_{5}} =\nu/D_5=1$,
respectively; see also
appendix~\ref{appendix_dimform}.
We probe different combinations of the chiral Mach number, using
${\rm Ma}_\mu=0.5$ and $2$, and the nonlinearity parameter
$\lambda_5=0.5$ and $8$, as given in the individual panels of figure~\ref{fig_timestep}.
In these examples, the chiral flipping rate and
the chemical potential have been neglected.
We note that this might be an incorrect simplification for proto-neutron
stars, where $\Gamma_\mathrm{f}$, being proportional to $m_e^2$, can reach
very large values \citep{GKR15,Dvornikov2017}.
DNS with non-vanishing $\Gamma_\mathrm{f}$ have been presented in
\cite{SRBBFRK17}, where it was shown that the evolution of $\mu_5$, and hence
$B_\mathrm{rms}$, can be strongly affected in the case without turbulence.
A detailed study of the effect of chiral flipping reactions in a turbulent
plasma and the potential damping out of
chiral magnetic instabilities, will be an interesting subject for future DNS.

In typical simulations of chiral MHD, the minimum time step is not determined
by the new contributions from the \texttt{chiral\_mhd.f90} module, but those can have
an important indirect effect caused by the amplification of the magnetic field.
It can be seen from figure~\ref{fig_timestep} that the chiral contributions,
indicated by blue colour, play mostly a subdominant role.
For comparison, time-step contributions from classical MHD are plotted,
including the acoustic time-step
$\delta t_c = c_{\delta t} \delta x/\mathrm{max} (c_\mathrm{s})$,
the advective time-step
$\delta t_{\UU} = c_{\delta t}\,\delta x/\max|\UU|$,
the Alfv\'en time-step
$\delta t_\mathrm{A} = c_{\delta t}\,\delta x/\max|v_\mathrm{A}|$,
and the resistive time-step
$\delta t_\eta =  c_{\delta t,\mathrm{v}}\,\delta x^2/\eta$,
where $c_{\delta t}$ and $c_{\delta t,\mathrm{v}}$ are user-defined
constants (the default values are $c_{\delta t}=0.9$ and
$c_{\delta t,\mathrm{v}}=0.25$),
and $v_\mathrm{A}$ is the Alfv\'en velocity.
From equations~(\ref{eq_dt1}) and (\ref{eq_dt2}), one could get the impression
that the chiral time step should become very small when the magnetic field
is strong, that is, at dynamo saturation.
However, $\BB^2$ occurs here always with a prefactor of $\lambda$ and, according
to equation~(\ref{eq_B2sat}), $\BB^2\lambda\approx\mu_{5,0}$.
The only regime where the chiral contribution to the time step becomes
important is the nonlinear phase of a plasma with large $v_\mu$ and low
$\lambda_5$; e.g.\ for $t \gamma_\mu>17$ in
figure~\ref{fig_timestep}(b).

Increasing ${\rm Ma}_\mu$ has an effect on the contributions to
the time step from classical MHD.
As mentioned before, a larger $\mu_{5,0}$ leads to a larger saturation magnetic
field strength; see equation~(\ref{eq_B2sat}).
This increases the Alfv\'en velocity and reduces the corresponding time step,
$\delta t_\mathrm{A}$; see the evolution of the black dotted lines in
figure~\ref{fig_timestep}.

\subsection{Minimum resolution}

The minimum resolution required for a simulation can be estimated
using the mesh Reynolds number, which is defined as
\begin{equation}
  \mbox{Re}_{\rm mesh}=\frac{\max(|\UU|) \delta x}{\nu},
\end{equation}
based on the resolution $\delta x$.
The value of $\mbox{Re}_{\rm mesh}$ should not exceed a certain value, which
is approximately $5$, but can be larger or smaller, depending on the nature of
the flow (smaller when the flow develops shocks, for example);
see the \pencilc\ manual, section~K.3.
Using this empirical value for a given viscosity (or resistivity) and given maximum
velocity, a minimum resolution $\delta x$ can be estimated.

In the following, velocities are given in units of
the speed of sound, $c_\mathrm{s}$.
Besides the sound speed, the turbulent velocity and shear velocities can occur
and determine $\mbox{Re}_{\rm mesh}$.
Most importantly at late stages of chiral dynamo simulations, i.e., in the
nonlinear dynamo phase and
especially close to saturation, the Alfv\'en velocity, $v_\mathrm{A}$, can
play a dominant role.
In dimensional units, $v_\mathrm{A, rms} = B_\mathrm{rms}/\sqrt{\meanrho}$, but
in code units with $\meanrho=1$, we have $v_\mathrm{A, rms} = B_\mathrm{rms}$.

In chiral MHD, the maximum $v_\mathrm{A}$ can be estimated from the conservation
law~(\ref{CL}).
Assuming that the magnetic field has a correlation length that is equal
to the size of the domain, the maximum magnetic field is of the order of
$(\mu_{5,0}/\lambda)^{1/2}$; see equation~(\ref{eq_B2sat}).
Hence, when a domain of size $(2\pi)^3$ is resolved with $N_\mathrm{grid}^3$
grid points,
we find the following requirement for the minimum resolution:
\begin{equation}
  N_\mathrm{grid} \gtrsim \left(\frac{\mu_{5,0}}{\lambda}\right)^{1/2} \frac{2\pi}{\nu \mbox{Re}_{\rm mesh, crit}}.
\end{equation}
One must not use too large values of \texttt{mu5\_const}
in \texttt{start.in} and too small values of \texttt{nu} (or \texttt{eta}) and
\texttt{lambda5} in \texttt{run.in}.
For example, when $\mu_{5,0}=10$, $\lambda=10^3$ and $\nu=10^{-3}$, a resolution
of more than $128^3$ mesh points is necessary.

We note that, in principle, larger saturation values of the magnetic field can
be calculated in the \pencilc\ by manually setting the value of $c_\mathrm{s}$
in \texttt{start.in} to a larger value, e.g.~$c_\mathrm{s}=2$.
This, however, is accompanied by a decrease of the simulation time step;
see the previous section.

\section{Numerical simulations in chiral MHD}
\label{sec_usage}

\subsection{The chiral MHD dynamo instability}

\subsubsection{Classical vs.\ chiral MHD}

\begin{figure}
\begin{center}
  \includegraphics[width=0.49\textwidth]{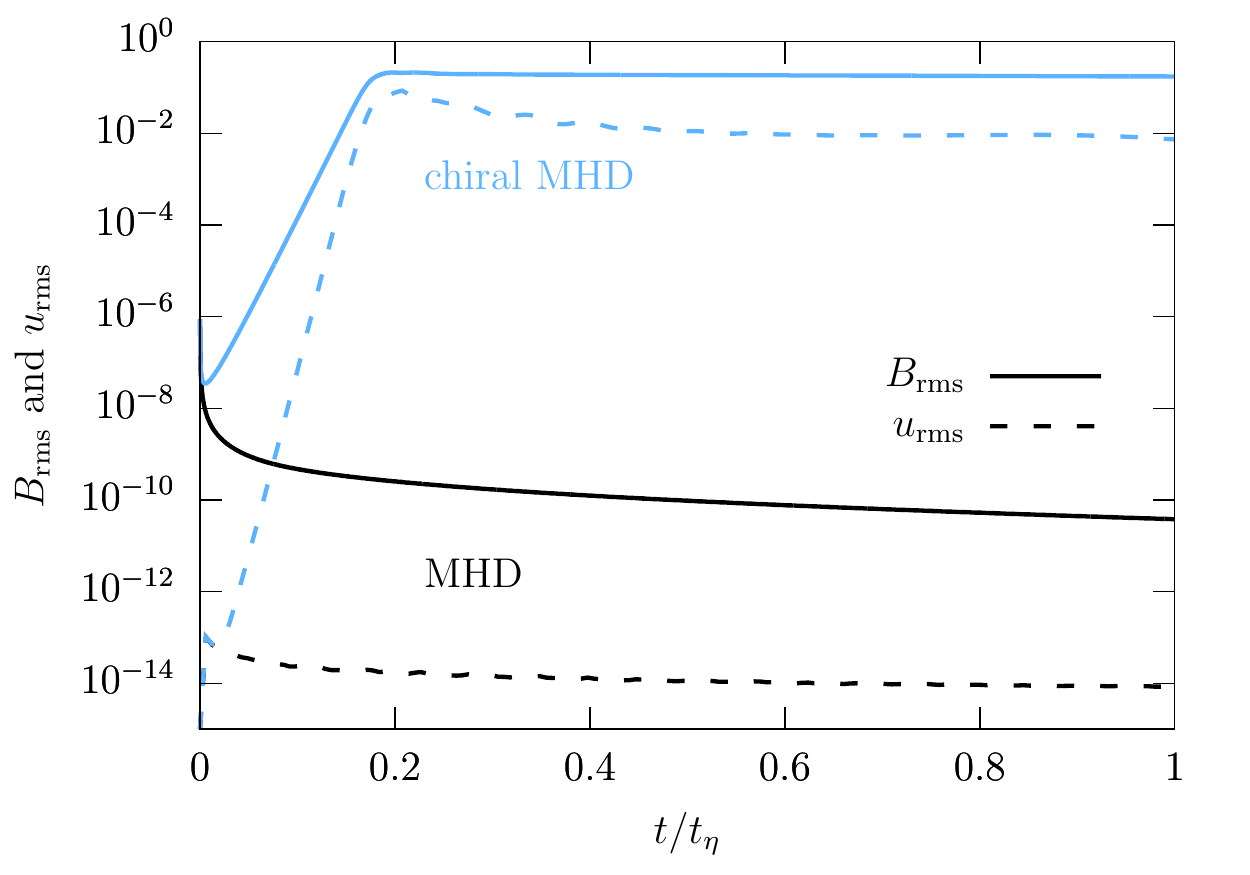}
\caption{Classical vs.\ chiral MHD. Two 2D simulations with a an
initially weak magnetic field and vanishing velocity field, without external
forcing of turbulence.
The black lines show the evolution of the rms magnetic fields strength
(solid black line) and the rms velocity (dashed black line) in the case of
classical MHD. Here both, $B_\mathrm{rms}$ and $u_\mathrm{rms}$, decay in time,
which is plotted in terms of the resistive time, $t_\eta\equiv{(\eta k_1)^{-1}}$.
When the chiral chemical potential is non-zero, as here shown with blue colour,
both, $B_\mathrm{rms}$ and $u_\mathrm{rms}$, grow exponentially over many
orders of magnitude, due to the laminar chiral MHD dynamo. (Colour online)}
\label{fig_classicalvschiral}
\end{center}
\end{figure}

The term $\nab\times(\eta\mu_5\BB)$ in the induction
equation~(\ref{ind-DNS}) drastically increases the range of
laminar and turbulent dynamos.
An example, where the evolution of a plasma with CME differs
strongly from the classical picture, is presented in figure~\ref{fig_classicalvschiral}.
The two 2D runs in domains of size $(2\pi)^2$ compared there, are resolved
by $256^2$ grid cells. They have periodic boundary conditions,
an initially vanishing velocity field, and a weak magnetic seed field.
In both cases, $\Pm=\mathrm{Pr}_5=1$, explicit viscosity, resistivity, and diffusivity
of $\mu_5$ have been included,
and the equation of state is that of an ideal gas.

The run presented as black lines in figure~\ref{fig_classicalvschiral} shows the
classical MHD case.
Here, as expected, the magnetic field decreases, since no classical dynamo
is operating in this system.
The blue lines in figure~\ref{fig_classicalvschiral} show the time evolution for a
typical chiral MHD scenario.
The simulation setup is chosen exactly in the same way as for the classical MHD
case, with the exception that the \texttt{chiral\_mhd.f90} module is activated,
that is, the induction equation includes the term $\nab\times(\eta\mu_5\BB)$
and equation~(\ref{mu5-DNS}) is solved to follow the evolution of $\mu_5$.
The simulation parameters are chosen such that ${\rm Ma}_\mu = 0.02$ and
$\lambda_5 = 0.002$.
The chiral instability scale is equal to $\mu_{5,0}/k_1=20$, where $k_1=1$ is the largest
wavenumber possible in the numerical domain.

The instability caused by the chiral term in the induction equation leads
to an increase of $B_\mathrm{rms}$ over more than 6 orders of magnitude
before saturation commences.
This occurs after less than $0.2$ diffusive times, $t_\eta$.
Simultaneously, the velocity $u_\mathrm{rms}$ increases by approximately 12
orders of magnitude due to driving of turbulence via the Lorentz force,
that is, via chiral-magnetic driving.

\subsubsection{Initial conditions for the chiral MHD dynamo}

Laminar dynamo theory predicts a scale-dependent growth rate of the magnetic
field according to equation~(\ref{eq_gammalam_max}).
If the initial magnetic field is distributed over all wavenumbers within the
box, like, for example, in case of Gaussian noise, the instability is strongest
on the scale $k_\mu=\mu_{5,0}/2$ and the rms magnetic field strength $B_\mathrm{rms}$
grows at the maximum rate $\gamma_\mu = \eta\mu_{5,0}^2/4$.
If the initial magnetic field is, however, concentrated at a single wavenumber $k_\mathrm{B}$,
which is the case for a force-free Beltrami field, e.g., for a vector potential
$\AAA = (\cos k_\mathrm{B} z,\;\sin k_\mathrm{B} z,\;0)$,
$B_\mathrm{rms}$ increases at the rate $\gamma(k_\mathrm{B})$.

\begin{figure}
\begin{center}
  \subfigure{\includegraphics[width=0.49\textwidth]{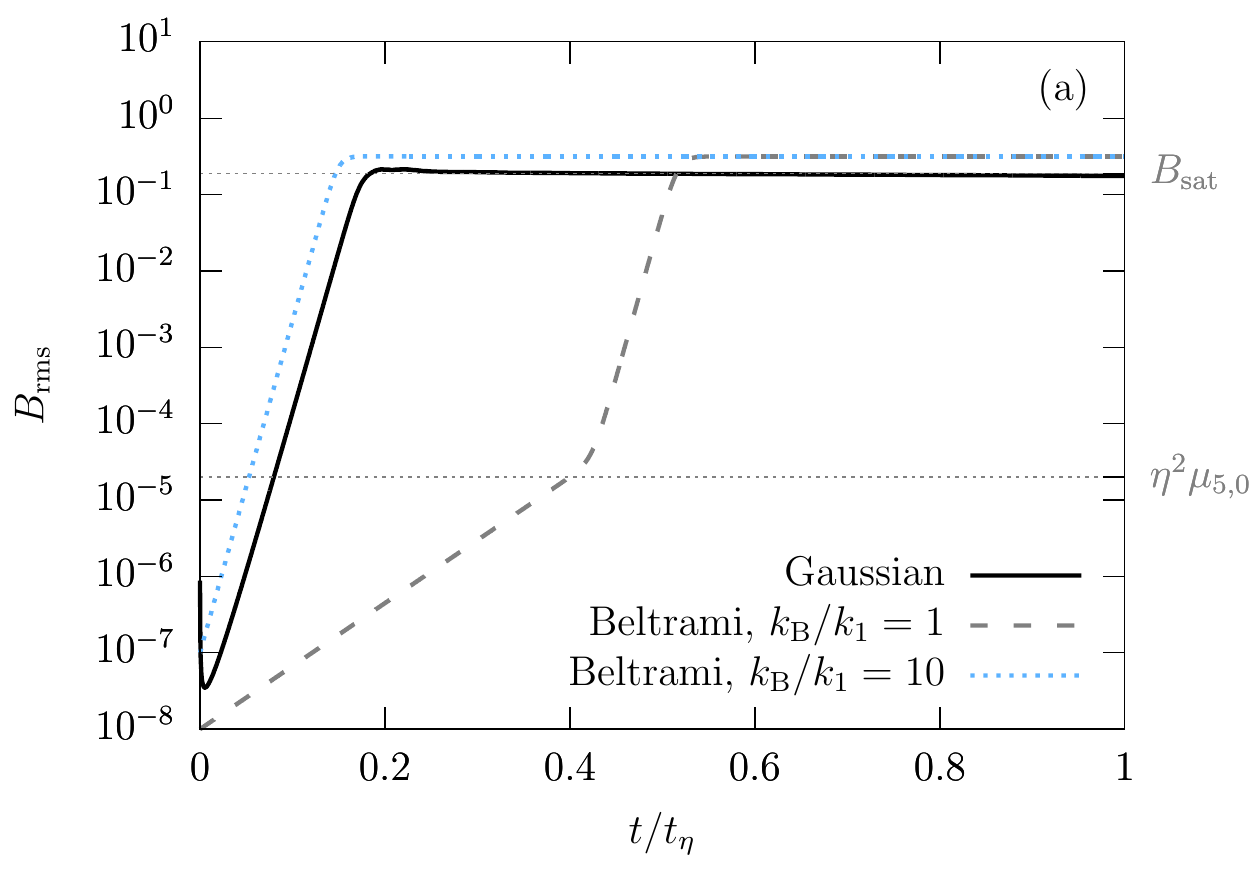}}
  \subfigure{\includegraphics[width=0.49\textwidth]{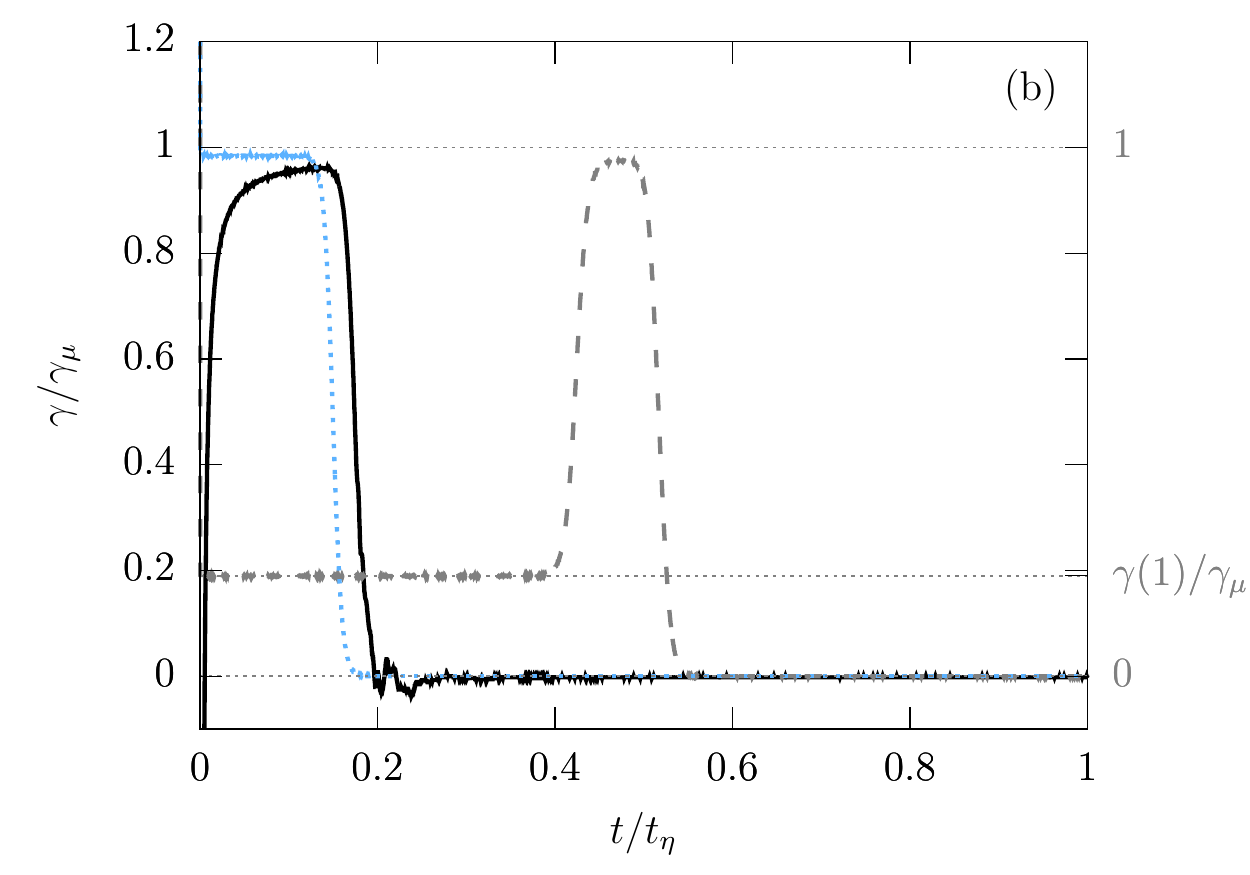}}
\caption{Simulations with different initial conditions of the magnetic field.
The 2D simulations with a domain of $(2\pi)^2$ and a resolution of $256^2$ have
parameters of $\Pm=1$, $\lambda_5=0.002$, and $k_\mu/k_1=10$.
The runs differ in the configurations of the initial magnetic field. We consider
Gaussian noise (black solid lines) and Beltrami fields with $k_\mathrm{B}/k_1=1$
(gray dashed lines) and $k_\mathrm{B}/k_1=k_\mu/k_1=10$ (blue dotted lines).
(a) Time evolution of the rms magnetic field strength.
(b) Time evolution of the growth rate of the magnetic field
strength. (Colour online)
}
\label{fig_initCond}
\end{center}
\end{figure}

A demonstration of the importance of the initial magnetic field configuration
is presented in figure~\ref{fig_initCond}, which shows the time evolution for
three 2D simulations.
All of these simulations have $k_\mu/k_1=10$.
The case with initial Gaussian noise increases with $\gamma_\mu$ until
saturation is reached at approximately $t=0.2~t_\eta$.
The Beltrami field, initiated at wavenumber $k_\mathrm{B}=1$, grows at
a rate $\gamma(1)<\gamma_\mu$.
Only once a field strength of $\eta^2\mu_{5,0}$ at $t\approx0.4~t_\eta$ is reached,
the field configuration has changed sufficiently such that the magnetic
energy is non-zero at $k_\mu$ and the $B_\mathrm{rms}$ continues to grow
with $\gamma_\mu$.

We note that a Beltrami initial field can also result in amplification with
$\gamma_\mu$, if it is concentrated around $k_\mathrm{B}=k_\mu$.
This is demonstrated by the simulation with $k_\mathrm{B}=k_\mu=20~k_1$,
which increases with the maximum possible growth rate from the beginning.
For all runs discussed above, the growth rates are shown in
figure~\ref{fig_initCond}(b) as a function of time.

\subsection{Chiral MHD in turbulence}

\subsubsection{Properties of chiral dynamos in chiral-magnetically and externally
driven turbulence}

The effects of turbulence on the evolution of a magnetic field in a chiral
plasma can be described by mean-field theory, which was reviewed briefly
in Section~\ref{sec_turbulentflows}.
The \pencilc\ allows for more detailed studies of chiral turbulent dynamos
without using simplifications of the equations that are made for an
analytical treatment.
Therefore, in the following we present two three-dimensional simulations in
domains of size $(2\pi)^3$ with periodic boundary conditions and a resolution
of $200^3$.
They are initiated with a weak random magnetic field and a chiral chemical
potential $\mu_{5,0} = 20 k_1$.
The two DNS are identical except for the fact that in one, turbulence is
driven externally at the wavenumber $k_{\rm f}=10\,k_1$.
We label the run with external forcing as ``$\mathrm{R}_\mathrm{f}$'', where ``f'' refers to
forcing, and the initially laminar
one as ``$\mathrm{R}_\chi$'', where ``$\chi$'' refers to chiral-magnetically driven
turbulence.

\begin{figure}[h!]
\begin{center}
  \subfigure{\includegraphics[width=0.45\textwidth]{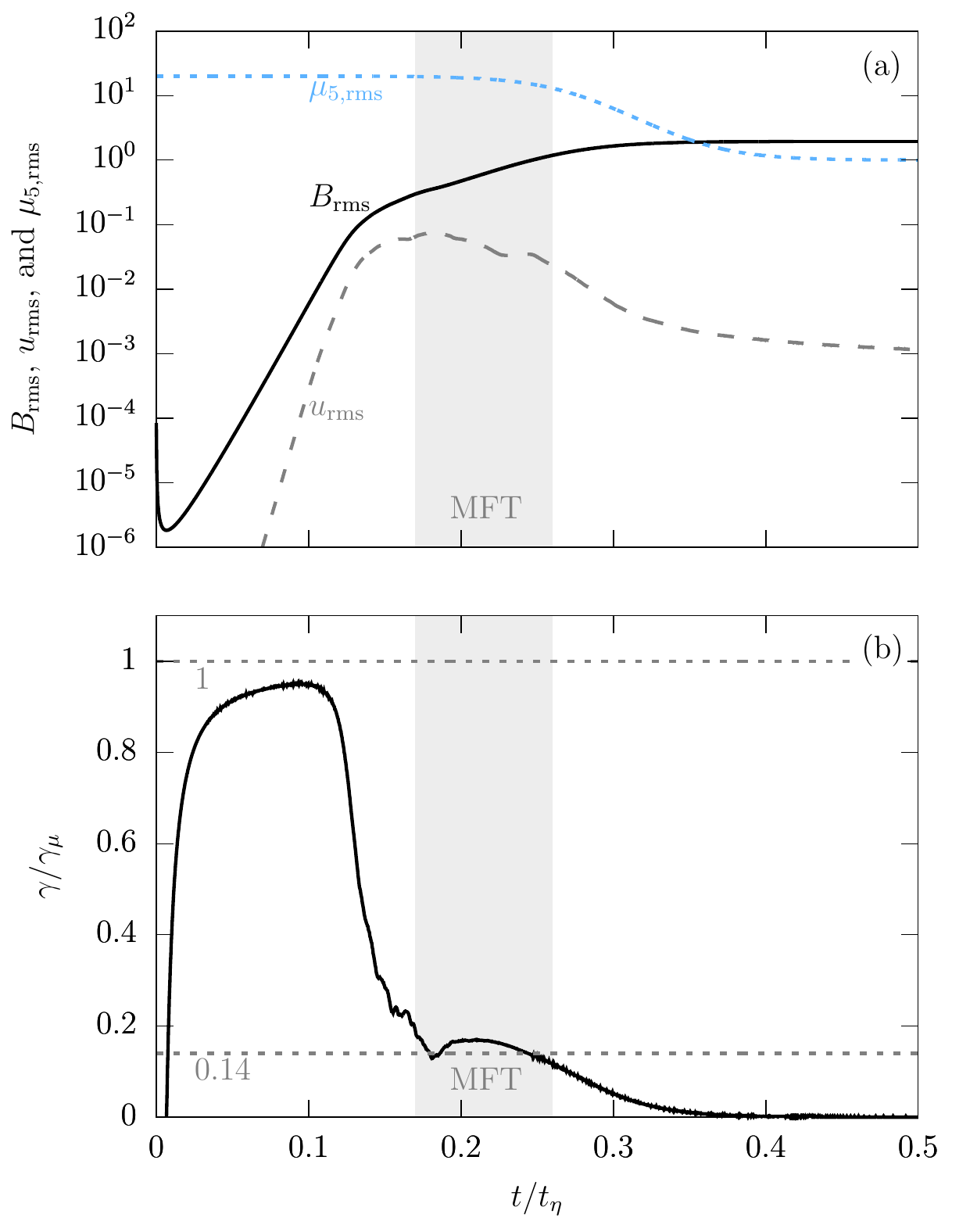}}
  \subfigure{\includegraphics[width=0.45\textwidth]{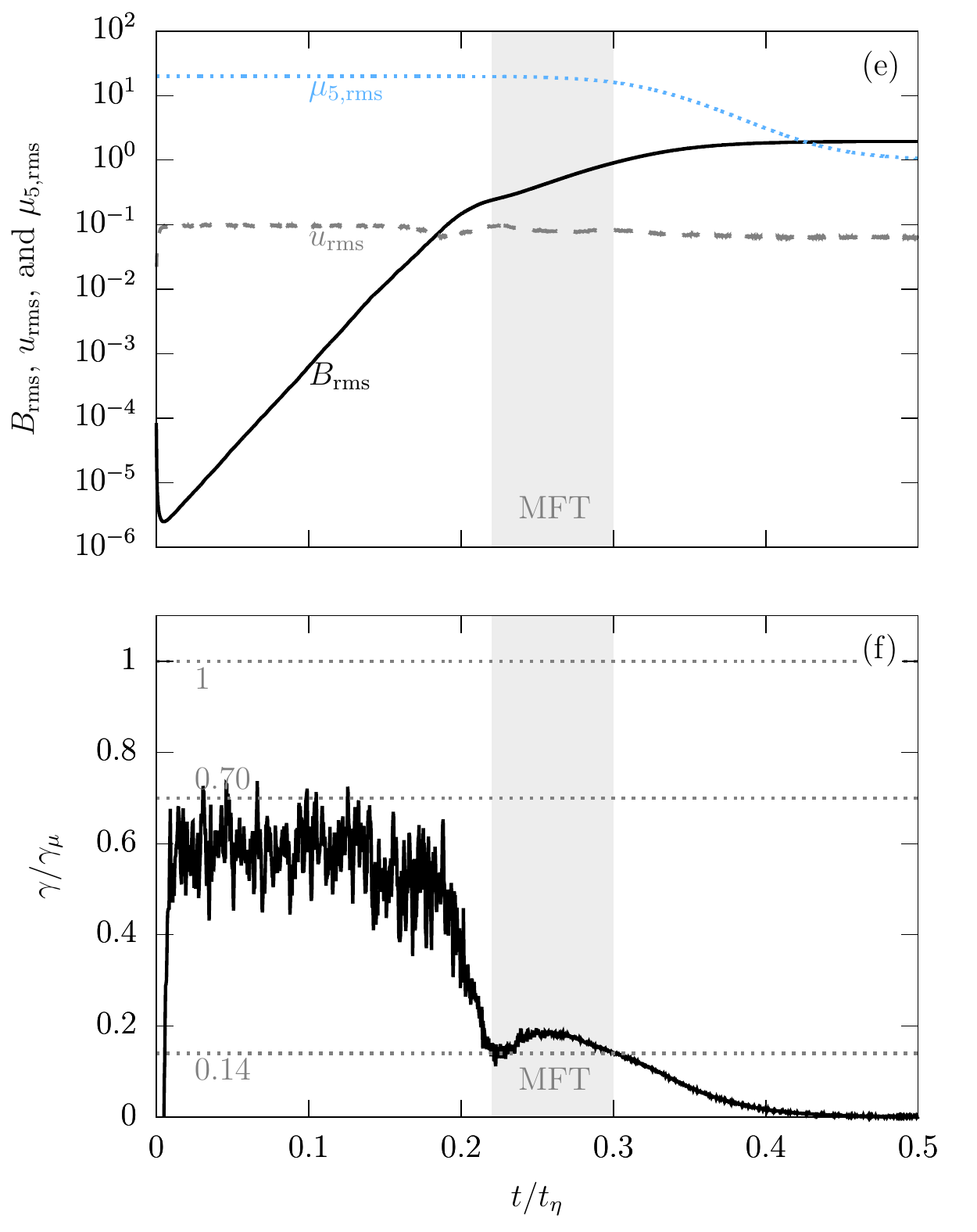}}
  \subfigure{\includegraphics[width=0.45\textwidth]{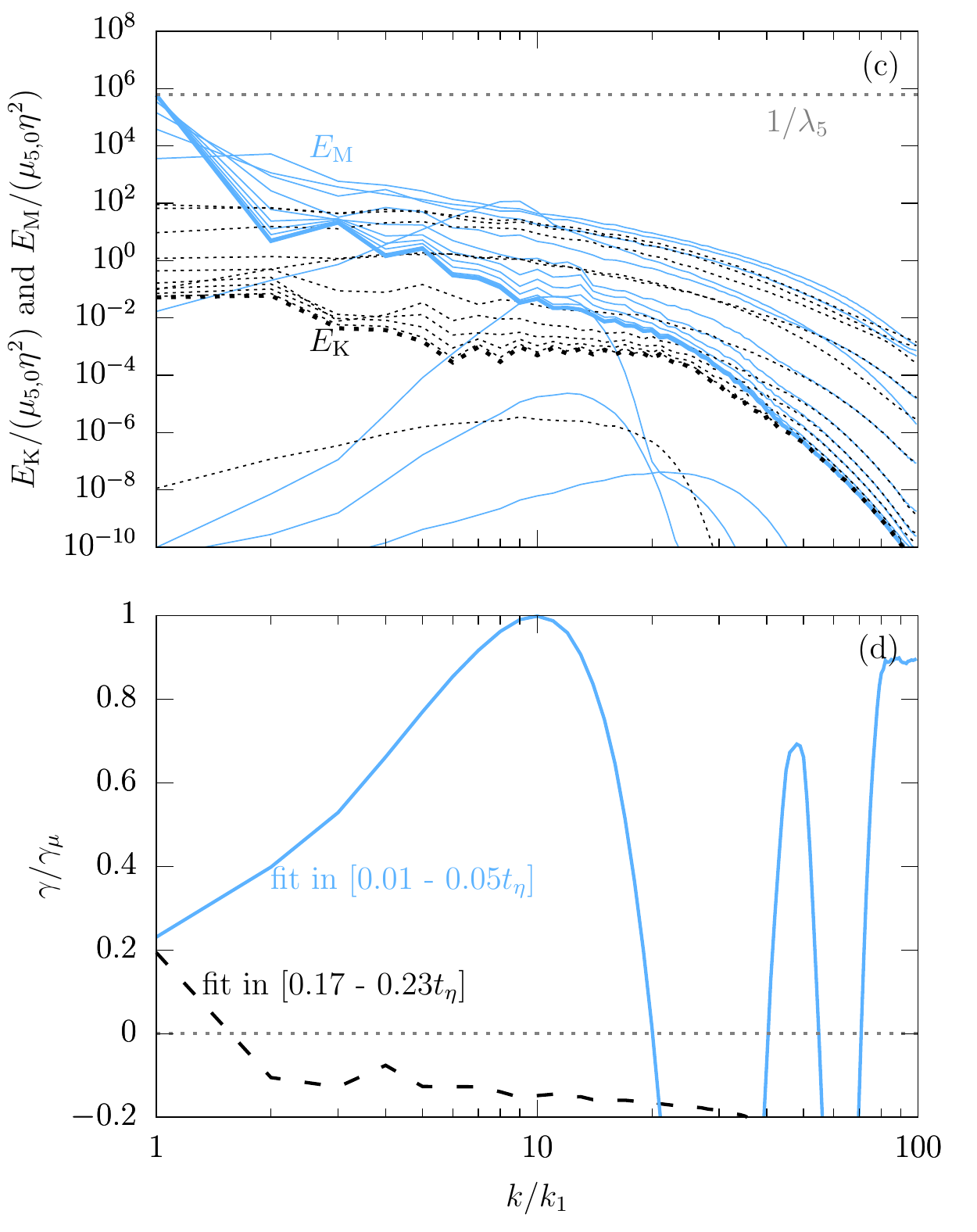}}
  \subfigure{\includegraphics[width=0.45\textwidth]{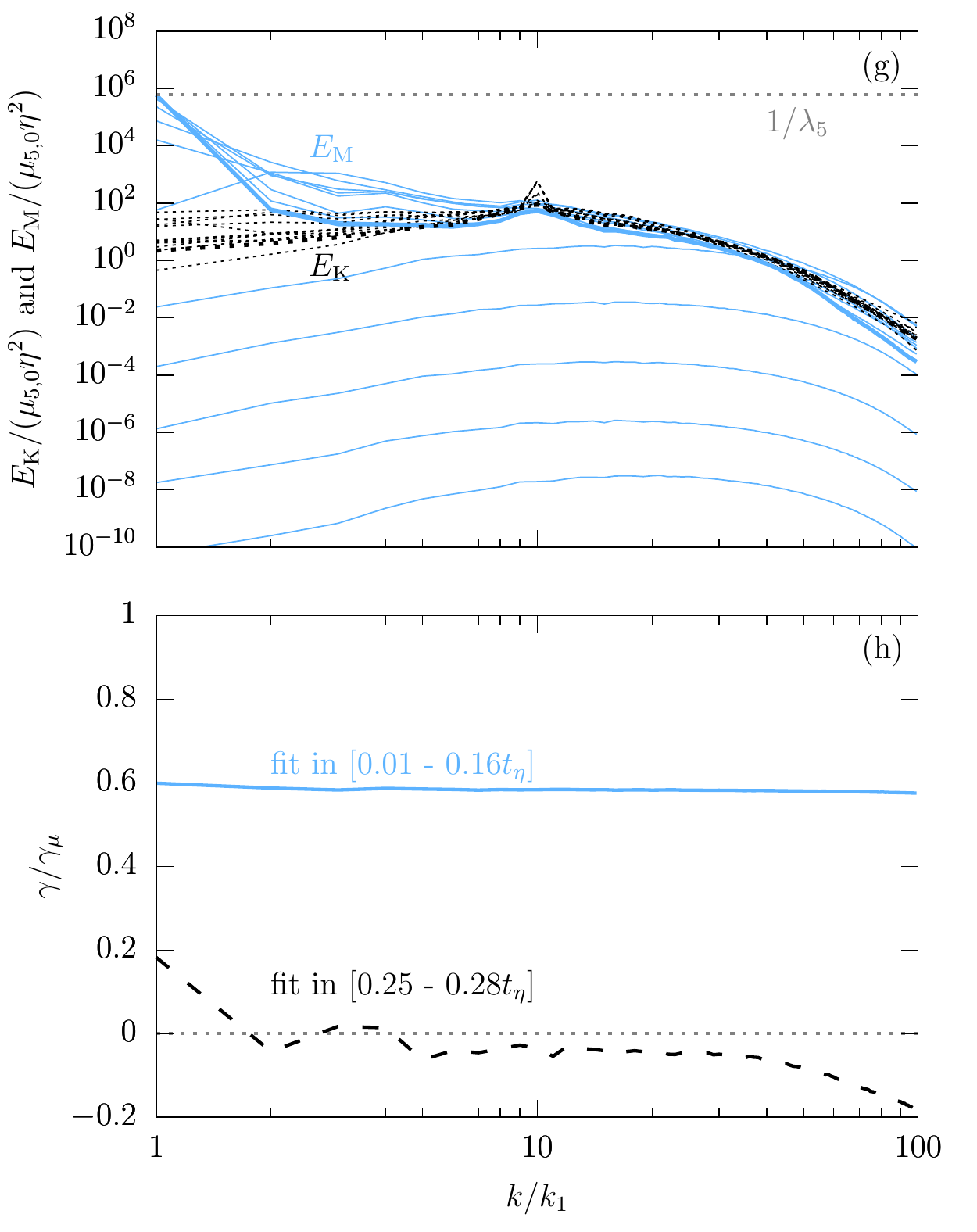}}
\caption{Direct comparison of chiral MHD dynamos in chiral-magnetically driven
turbulence (left) and externally forced turbulence (right).
In the first two rows, we present the time series of $B_\mathrm{rms}$,
$u_\mathrm{rms}$, $\mu_{5,\mathrm{rms}}$, and $\gamma/\gamma_\mu$.
The time interval in which linear mean-field theory (``MFT'') applies
is highlighted by a gray background.
In the third row, the dotted black lines show kinetic energy spectra
and the solid blue lines show magnetic energy spectra in
uniform time intervals.
The final spectra, obtained at $t=0.5t_\eta$ are plotted as thick lines.
The scale
dependence of the growth rate at different time intervals is shown in the last
line. (Colour online)
}
\label{fig_turbulence}
\end{center}
\end{figure}

In figure~\ref{fig_turbulence}, the two simulations are compared directly, where
we present run~$\mathrm{R}_\chi$ in the left panels and run~$\mathrm{R}_\mathrm{f}$ in the right panels.
The main differences between the two cases are clearly visible in the upper
panels, which show the time evolution of $B_\mathrm{rms}$, $u_\mathrm{rms}$,
and $\mu_{5,\mathrm{rms}}$.
The magnetic field grows much faster in run~$\mathrm{R}_\chi$,
which can be seen more clearly in the second row of figure~\ref{fig_turbulence},
where the evolution of $\gamma$, normalised by the laminar growth rate $\gamma_\mu$,
is presented.
Simultaneously with $B_\mathrm{rms}$ growing at a rate of $\gamma_\mu$, $u_\mathrm{rms}$ grows at a rate of
approximately $2\gamma_\mu$, as expected for driving through the Lorentz force.
Once the kinetic energy becomes comparable to the magnetic energy at
$t\approx 0.12~t_\eta$, $\gamma$ decreases as a result of additional
turbulent diffusion.
When turbulence is forced externally, we observe an initial amplification of
$B_\mathrm{rms}$ with a growth rate that is reduced as compared to $\gamma_\mu$.
As in run~$\mathrm{R}_\chi$, mean-field effects,
e.g.\ turbulent diffusion and the $\alpha_\mu$ effect, occur in $\mathrm{R}_\mathrm{f}$ once
$B_\mathrm{rms}\approx u_\mathrm{rms}$, resulting in an overall decrease of the growth
rate at $t\approx 0.2~t_\eta$.

A major difference between externally and chiral-magnetically driven turbulence
appears in the comparison of the energy spectra; see the third row of
figure~\ref{fig_turbulence}.
While in the initially laminar run $\mathrm{R}_\chi$, the magnetic field instability
occurs at wavenumber $k_\mu=\mu_{5,0}/2=10$, we observe a scale-independent growth of the
magnetic energy in $\mathrm{R}_\mathrm{f}$.
The growth rate is presented as a function of $k$ in the bottom panel.
This dependence is clearly different from the parabola shape predicted
from theory, see equation~(\ref{eq_gammalam}), and is the result of mode
coupling.
Hence, in the presence of turbulence, the magnetic field grows at a reduced rate,
which can be estimated \\

\vspace{1cm}
\newpage

\noindent
as
\begin{align}
  \tilde\gamma(\mu_{5,0}) =  \frac{1}{\mu_{5,0} - k_1} \int_{k_1}^{\mu_{5,0}} \gamma(k)~\mathrm{d}k
                       =\,& \frac{1}{6} \eta (\mu_{5,0}-k_1) (\mu_{5,0}+2 k_1)\nonumber\\
                       =\,&  \frac{2}{3} \frac{(\mu_{5,0}-k_1) (\mu_{5,0}+2 k_1)}{\mu_{5,0}^2} \gamma_\mu. 
\end{align}
The value $\tilde\gamma(\mu_{5,0})$ reaches its maximum
of $(3/4) \gamma_\mu$ at $\mu_{5,0}=4 k_1$.
When $\mu_{5,0}$ is increased, the initial growth rate of the magnetic field
decreases; e.g.\ for $\mu_{5,0} = 20 k_1$ we find
$\tilde\gamma(20 k_1) \approx 0.70 \gamma_\mu$, as expected for our DNS,
and for $\mu_{5,0} = 100 k_1$ we estimate
$\tilde\gamma(100 k_1) \approx 0.67 \gamma_\mu$.
The growth rate of $0.70 \gamma_\mu$ is indicated as a horizontal dotted
line in figure~\ref{fig_turbulence}(f).

In runs~$\mathrm{R}_\chi$ and $\mathrm{R}_\mathrm{f}$, the presence of an $\alpha_\mu$
effect, which drives a large-scale dynamo, can only be seen at late times,
shortly before dynamo saturation.
As discussed in \citet{SRBBFRK17} and in the following section in more detail, the growth rates
measured in DNS at late times agree approximately with the theoretical prediction from
equation~(\ref{eq_gammaturb_max}).

The DNS results suggest that mean-field effects in the evolution of the
magnetic field occur once the magnetic energy is larger than the kinetic energy.
In terms of normalised quantities, this translates to
$B_\mathrm{rms}>u_\mathrm{rms}$.
Whether or not the system can reach this condition is determined by the
chiral conservation law and, in particular, by the value of $\lambda$.
Using equation~(\ref{eq_B2sat}), one finds that
mean-field effects in the evolution of $B_\mathrm{rms}$ occur for
$\lambda\lesssim \mu_{5,0}/(\xi_\mathrm{M} u_\mathrm{rms}^2)$.

\subsubsection{Indirect evidence for the $\alpha_\mu$ effect}

In the limit of large $\Rm$ and a weak mean magnetic field, the theoretically
expected growth rate~(\ref{eq_gammaturb_max}) can be written as
\begin{eqnarray}
  \gamma_\alpha(\Rm) = \gamma_\mu \frac{\left[1 - (2/3)\, \log \, \Rm \right]^2}{1 + \, \Rm/3}.
\label{eq_gammaturb_max2}
\end{eqnarray}
In fact, at $\Rm\approx4.5$, $\gamma_\alpha(\Rm)$ as given above vanishes.
However, at these moderate values of $\Rm$,
deviations from the $\tau$ approximation, used for deriving equation~(\ref{eq_gammaturb_max2}), can be expected.
The maximum of the dynamo growth rate in turbulence is expected for
$\Rm\approx38.7$, where $\gamma_\alpha(38.7)\approx 0.15\gamma_\mu$.

In the DNS presented in figure~\ref{fig_turbulence}, the maximum Reynolds
number for run $\mathrm{R}_\chi$ is $\approx 25$, which is comparable to
the value of $\Rm$ achieved in run $\mathrm{R}_\mathrm{f}$ via external forcing.
Based on mean-field theory we expect $\gamma_\alpha(25) \approx 0.14 \gamma_\mu$,
which is indicated as horizontal dotted lines in figures~\ref{fig_turbulence}(b)
and~\ref{fig_turbulence}(f).
The agreement between the growth rate measured in DNS and mean-field theory,
shown in these two examples,
can be viewed as indirect evidence for the existence of the $\alpha_\mu$ effect.

\subsection{MHD waves and the CME}
\label{sec_MHDwaves}

The chiral asymmetry also affects the dispersion relation for
MHD waves in a plasma.
If a chiral instability is excited, it has a direct effect on the
the frequencies and amplitudes of Alfv\'en and magnetosonic waves
through the amplification of the magnetic field.
How the dispersion relation in chiral MHD differs from the
one in classical MHD has been shown in \citep{REtAl17}.
The general expression for a compressible flow is given by
\begin{eqnarray}
  \left(\omega^2 - \oA^2\right) \, \Big[\omega^4 - \omega^2 \, ({\bm v}_A^2 + c_{\rm s}^2) k^2 + \oA^2  c_{\rm s}^2 k^2 \Big]
  - \omega^2 \,  (v_\mu \, k)^2 \, \left(\omega^2 - c_{\rm s}^2 k^2\right)=0,
\label{eq_disrel}
\end{eqnarray}
where $\oA ={\bm k} \cdot {\bm v}_A$ is the frequency of Alfv\'{e}n
waves in the absence of the CME.
The complexity of the dispersion relation~(\ref{eq_disrel}) indicates that in
chiral MHD,
the Alfv\'en and magnetosonic waves are strongly affected by a non-zero $\mu_5$.
For solutions of equation~(\ref{eq_disrel}) as a function
of the angle between the wavevector ${\bm k}$ and the
background magnetic field, we refer to figure~1 of \citet{REtAl17}.
In summary, the frequencies of the Alfv\'en wave and the magnetosonic wave
are increased for a weak magnetic field, while the frequency of the slow
magnetosonic wave is decreased in chiral MHD.

We use the \pencilc\ to study the properties of MHD waves in chiral MHD.
To this end,
we set up 1D simulations with an imposed magnetic field of the form
$(B_0,0,0)$ and $B_0=0.1$.
As initial condition for the magnetic field, we use \texttt{Alfven-x}, which
creates an Alfv\'en wave travelling in the $x$ direction:
\begin{eqnarray}
   A_z \propto \sin(k_x x -\oA t).
\label{eq_Alfven1}
\end{eqnarray}
The simulations of a domain with an extension $[-\pi,\pi]$ are resolved by 
$128$ grid points and the magnetic Prandtl
number is $1$.

\begin{figure}
\begin{center}
  \subfigure{\includegraphics[width=0.32\textwidth]{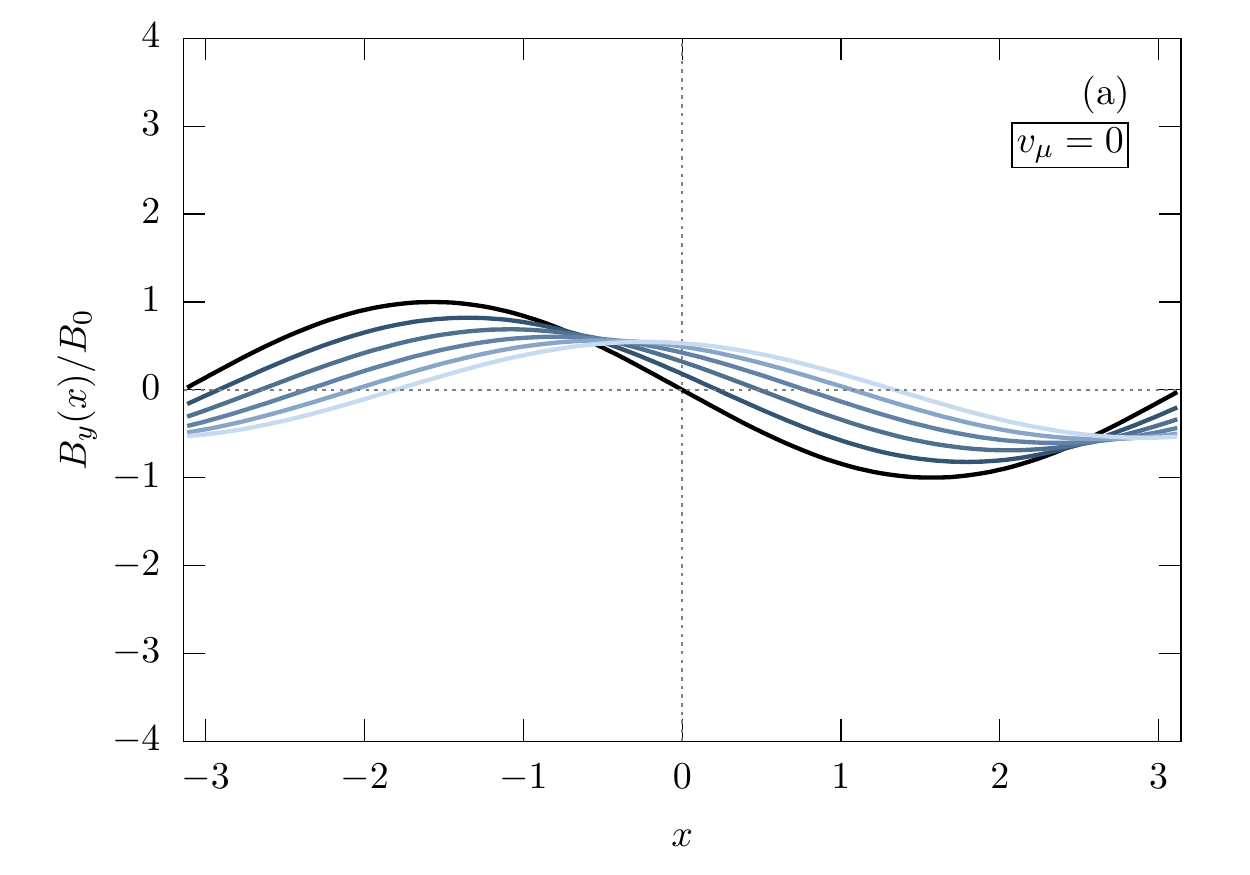}}
  \subfigure{\includegraphics[width=0.32\textwidth]{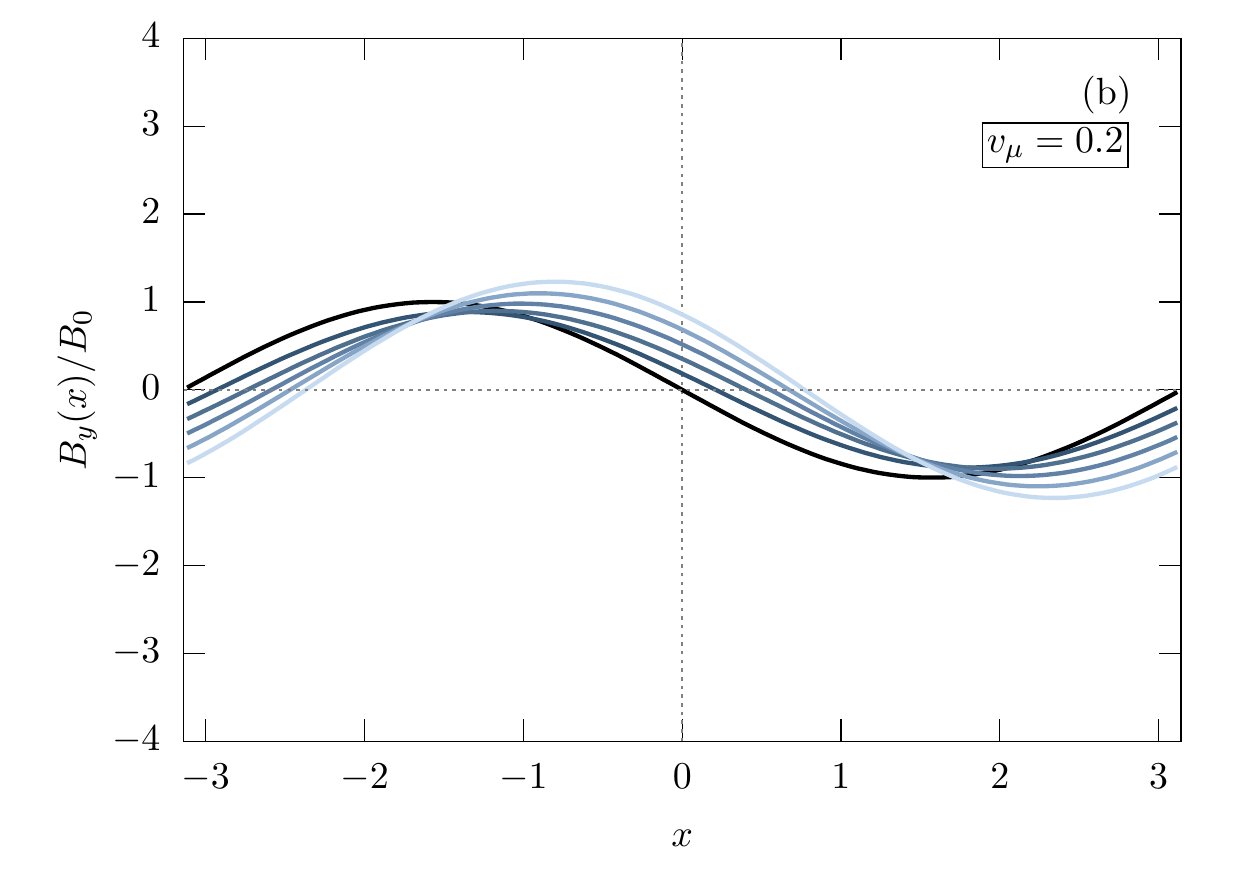}}
  \subfigure{\includegraphics[width=0.32\textwidth]{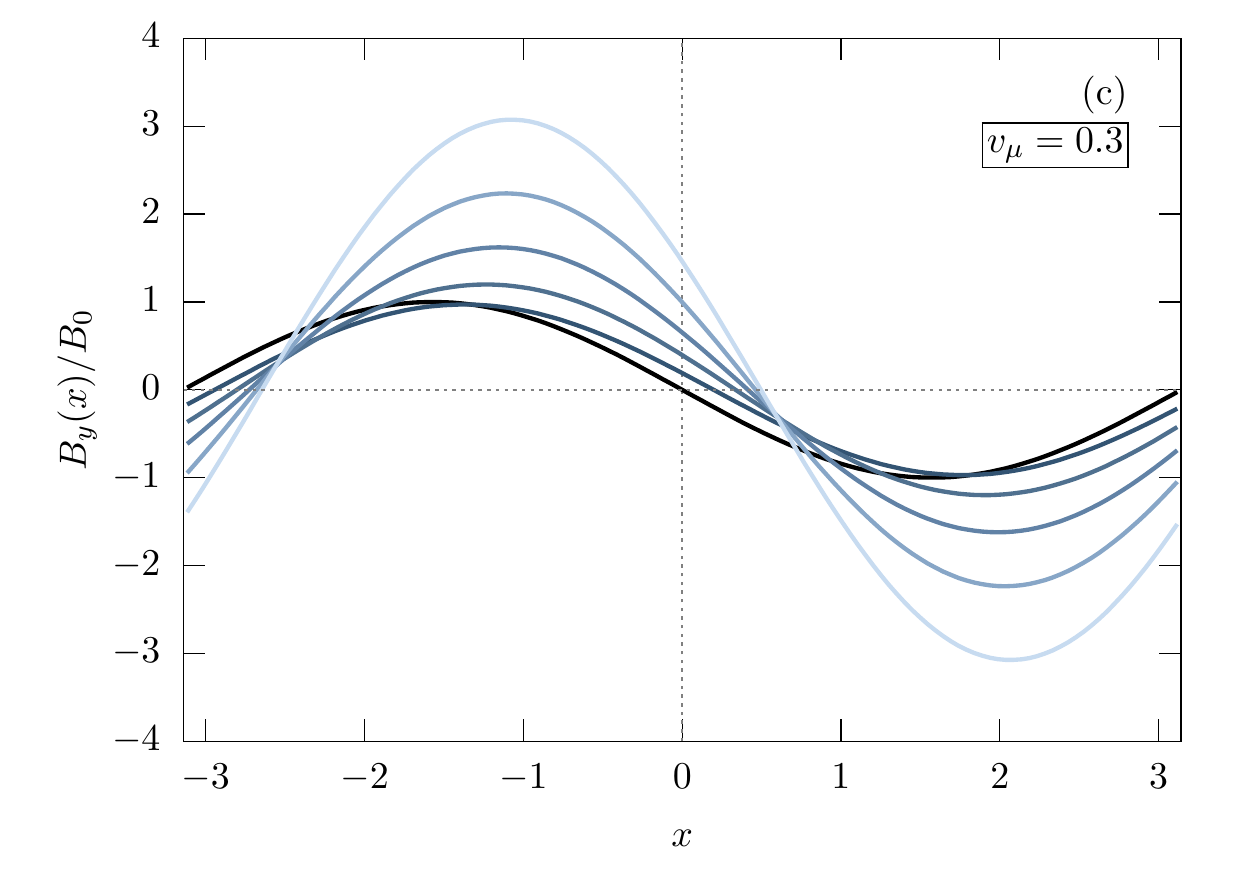}}
\caption{Propagation of a one-dimensional Alfv\'en wave for different
values of $\boldsymbol{v_\mu}$.
From left to right, $v_\mu$ increases from $0$ (classical MHD) to $v_\mu=0.3$.
The black lines show the initial condition of $B_y$ ($t=0$) and later times are
indicated by blue colour with the lightest blue indicating the last time shown
in the figure ($t=0.16~t_\mathrm{P}$ with $t_\mathrm{P}$ being the period
of a classical Alfv\'en wave).
The time difference between neighbouring lines is constant and the magnetic field
is normalised to its initial amplitude. (Colour online)}
\label{fig_By_x}
\end{center}
\end{figure}

In figure~\ref{fig_By_x}, the effect of changing $v_\mu$ on the propagation
of the wave is illustrated.
In the left panel, the classical MHD case with $v_\mu= 0$ is shown for reference.
Here the Alfv\'en wave is damped, leading to a decrease of the amplitude in
time and a propagation of the peak to the right with the Alfv\'en velocity
$v_\mathrm{A}$.
In the middle panel, a chiral MHD run is shown with $v_\mu=0.2$ and in the
right panel with $v_\mu=0.3$.
In all panels of figure~\ref{fig_By_x}, we show the same time interval, and
curves of the same colour indicate the same times.
One clearly sees in these simulations that, in chiral MHD,
the wave propagates more slowly, while its amplitude increases
due to the chiral dynamo instability.

\begin{figure}
\begin{center}
  \subfigure{\includegraphics[width=0.49\textwidth]{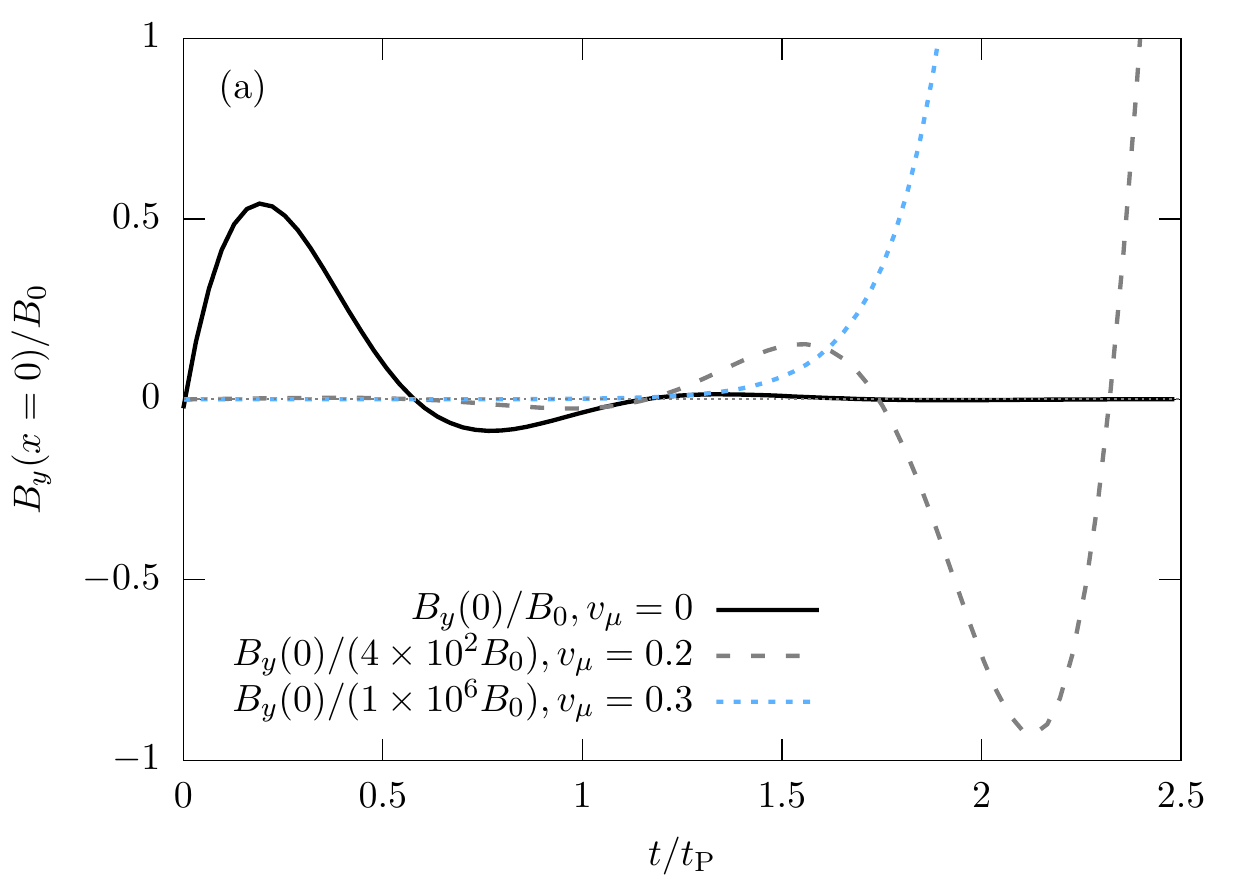}}
  \subfigure{\includegraphics[width=0.49\textwidth]{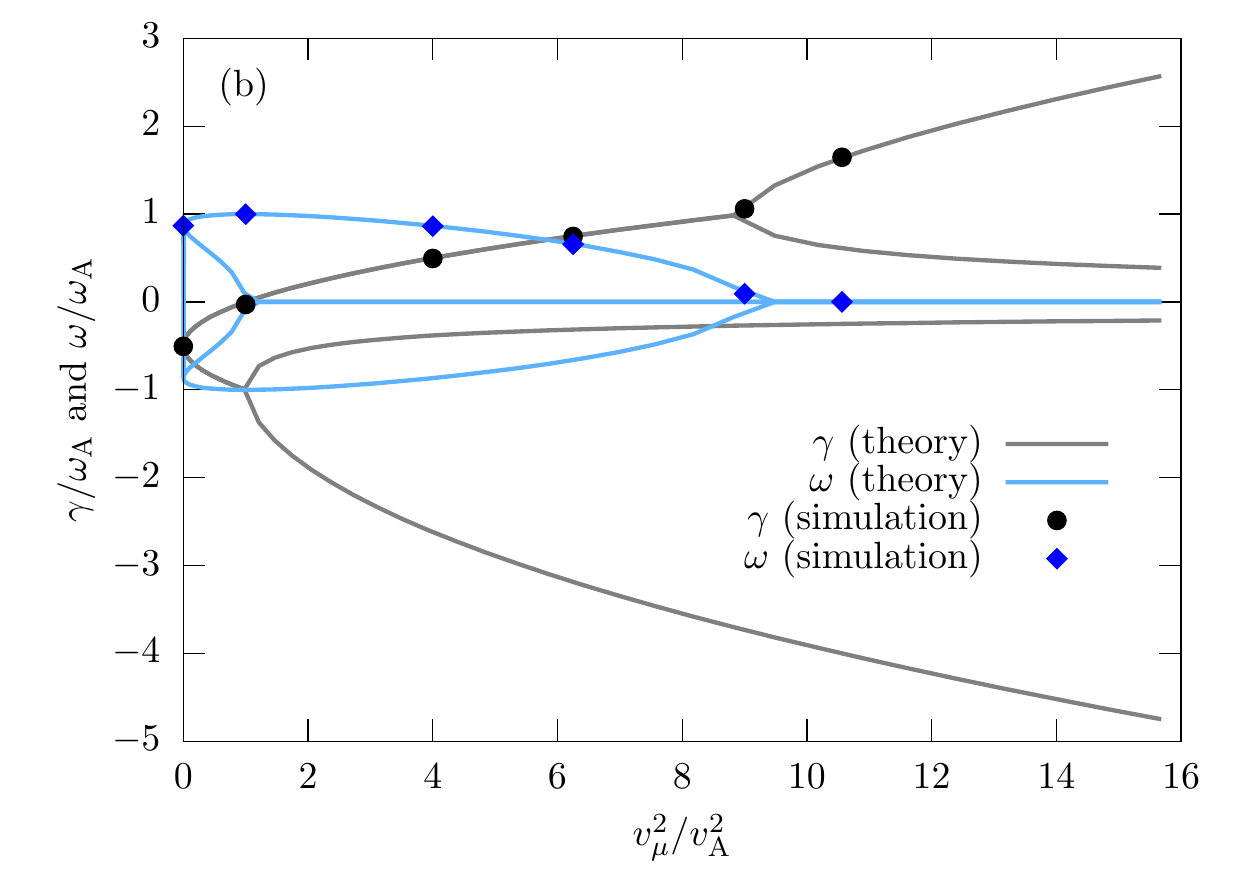}}
\caption{Effect of the CME on a one-dimensional Alfv\'en wave for different
values of $\boldsymbol{v_\mu}$.
(a)
The displacement of $B_y/B_0$ in the middle of the one-dimensional domain
(at $x=0$) for the three runs presented in figure~\ref{fig_By_x}.
The black solid curve, with $v_\mu=0$, shows the classical damped Alfv\'en wave
with the normal Alfv\'en frequency $\omega_\mathrm{A}$.
For better visibility, we present the case of $v_\mu=0.2$ (gray dashed line)
divided by a factor of $4\times10^2$ and the case of $v_\mu=0.3$ (blue dotted
line) divided by a factor of $10^6$.
(b)
Solution of the dispersion relation~(\ref{eq_disrel}) and comparison with simulation
results for runs with different $v_\mu$.
Theoretical solutions for the growth rate $\gamma$ (real parts of solutions)
are shown as gray lines and solution for the frequency $\omega$ (imaginary
parts of solutions) as light blue lines.
On top of these curves results from DNS with different values of $v_\mu$ are
presented: Black dots show the measured growth rates and blue diamonds the
frequencies, both of which are obtained from fits to $B_y(x=0,t)$.
(Colour online)}
\label{fig_By0_disrel}
\end{center}
\end{figure}

In figure~\ref{fig_By0_disrel}(a), we present the time evolution
of $B_y(x=0)$ for all runs shown in figure~\ref{fig_By_x}.
In the classical MHD case, the Alfv\'en wave is damped and the amplitude becomes
indistinguishable from zero after approximately 1.5 periods.
The case of $v_\mu=0.2$ corresponds to a wave with a growing amplitude.
By contrast,
for $v_\mu=0.3$ we observe a non-oscillating solution with exponentially growing amplitude.
By fitting $B_y(x=0)$, we can obtain the growth rate $\gamma$ and the
frequency $\omega$.
The results of these fits are presented in figure~\ref{fig_By0_disrel}(b)
as a function of $(v_\mathrm{A}/v_\mu)^2$.
The values measured in DNS agree well with the solutions of the dispersion
relation~(\ref{eq_disrel}), which are presented as solid lines.

\subsection{The role of the chemical potential}

In all the simulations discussed up to now, the ordinary chemical potential $\mu$ has been neglected.
For finite $C_5$, however, the evolution of $\mu_5$ is coupled to that of $\mu$ via
the term $- C_5 ({\BB} {\bm \cdot} \nab) \mu$, which could potentially
affect chiral MHD dynamos and excite collective modes.
In this section we present
DNS of chiral MHD including a non-zero chemical potential.

\subsubsection{Effects on the nonlinear evolution of the chiral dynamos}

\begin{figure}
\begin{center}
  \subfigure{\includegraphics[width=0.49\textwidth]{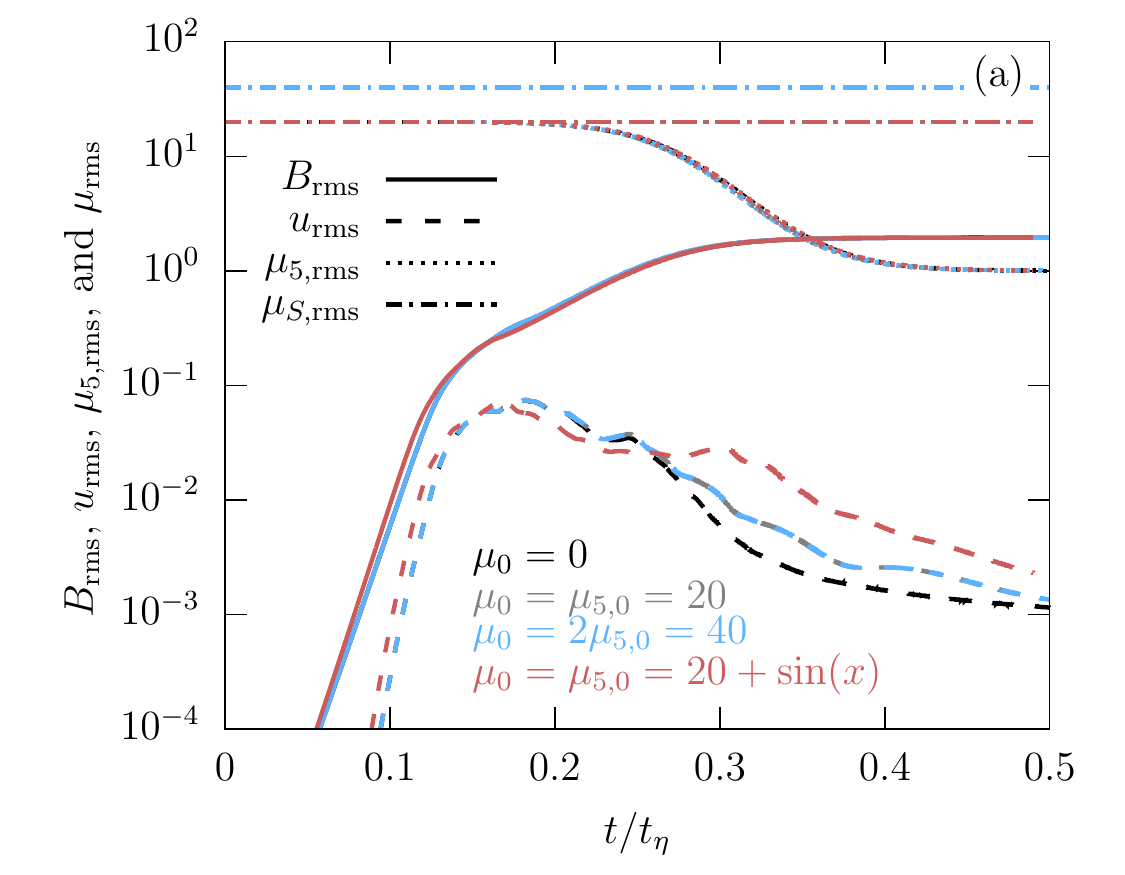}}
  \subfigure{\includegraphics[width=0.49\textwidth]{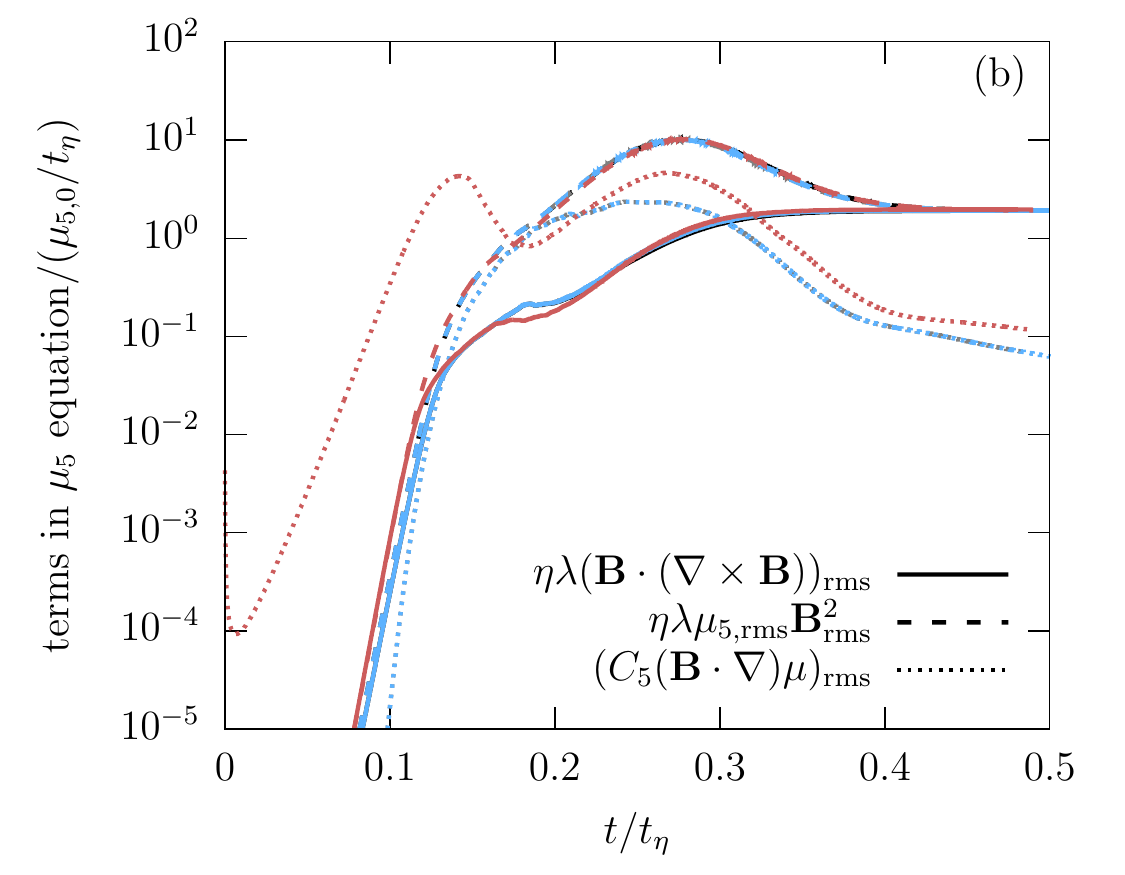}}
\caption{Chiral MHD dynamos with different chemical potentials $\boldsymbol{\mu}$.
The figure shows results for various initial conditions with
$\mu_{0}=0$ and $\mu_{5,0}=20$ (black lines),
$\mu_{0}=\mu_{5,0}=20$ (grey lines),
$\mu_{0}=2\mu_{5,0}=40$ (blue lines), and
$\mu_{0}=\mu_{5,0}=20+\sin(x)$ (red lines).
(a) Time evolution of $B_\mathrm{rms}$, $u_\mathrm{rms}$,
$\mu_{5,\mathrm{rms}}$, and $\mu_\mathrm{rms}$, as indicated in the label.
(b) Time evolution of the terms in the evolution equation of $\mu_5$,
normalised to $\mu_{5,0}/t_\eta$. (Colour online)}
\label{fig_muS}
\end{center}
\end{figure}

To explore the role of the chemical potential in chiral dynamos,
we repeat our exemplary run for
chiral-magnetically driven turbulence, which is presented in
figures~\ref{fig_turbulence}(a)--(d).
Here we solve the full system of equations~(\ref{ind-DNS})--(\ref{mu-DNS}),
neglecting only the forcing term in the Navier-Stokes equation and chirality
flipping in the evolution equation of $\mu_5$.
The diffusivity $D_\mu$ has the same value as $D_5=\eta=\nu$ and
for the coupling constants in equations~(\ref{mu5-DNS}) and (\ref{mu-DNS})
we use $C_5=C_\mu=1$, respectively.
Three different initial conditions for the chemical potential are considered:
$\mu_0=\mu_{5,0}=20 k_1$, which illustrates the case with only left- or
right-handed fermions, a case with $\mu_0=2\mu_{5,0}=40 k_1$, and
$\mu_0=\mu_{5,0}= (20 +\sin(x))k_1$.

In figure~\ref{fig_muS}(a), these three runs are compared with the simulation
presented in figures~\ref{fig_turbulence}(a)--(d), where $\mu_0=0$ and the
evolution of $\mu$ has been neglected.
Different colours in figure~\ref{fig_muS} indicate results for different runs.
Black lines show the case presented in figure~\ref{fig_turbulence}(a)--(d),
grey lines the case where $\mu_0=\mu_{5,0}$, blue lines the
case where $\mu_0=2\mu_{5,0}$, and red lines the case with
a sinusoidal spatial variation in $\mu_0$ and $\mu_{5,0}$.
As one may expect, only minor differences in the nonlinear phase
of $u_\mathrm{rms}$ and $\mu_{5,\mathrm{rms}}$ can be noticed between the different runs.
Naturally, the small deviations of $u_\mathrm{rms}$ and
$\mu_{5,\mathrm{rms}}$ do not depend on the value of $\mu_0$ if it is
constant and non-zero.
A slightly larger change in the non-linear evolution of $u_\mathrm{rms}$
as compared to $\mu_0=0$, is seen in the case of an initial sinusoidal variation of
$\mu_0$, due to the larger gradients in $\mu$.

For a better understanding of the evolution of $\mu_5$ in the different DNS,
we present in figure~\ref{fig_muS}(b) the time evolution of the various terms
in equation for $\mu_5$:
$\lambda \, \eta \, {\BB} {\bm \cdot} (\nab \times {\BB})$,
$\lambda \, \eta \, \mu_5 {\BB}^2$, and $C_5 ({\BB} {\bm \cdot} \nab) \mu$.
All of these terms are normalised by $\mu_{5,0}/t_\eta$ and the same colour
code is used as in figure~\ref{fig_muS}(a).
It is important to note that the first two of these terms are only relevant
for the nonlinear dynamo phase, e.g.\ when $\BB$ is large.
As can be seen in the plot, the term $\lambda \, \eta \, \mu_5 {\BB}^2$ is
eventually responsible for decreasing $\mu_5$ and therefore shutting off the
dynamo.
We observe only very minor differences between all three runs in the terms
$\lambda \, \eta \, {\BB} {\bm \cdot} (\nab \times {\BB})$ and
$\lambda \, \eta \, \mu_5 {\BB}^2$.
Obviously, the term $C_5 ({\BB} {\bm \cdot} \nab) \mu$ evolves very differently
for a constant $\mu_0$ and one with a sinusoidal variation.
Therefore the red dotted line is initially dominant.
At time $\approx 0.12 t_\eta$, it drops and the non-linear dynamo phase in this case
becomes comparable to the cases with constant $\mu_0$.
For extreme gradients in $\mu$, the term $C_5 ({\BB} {\bm \cdot} \nab) \mu$ could,
in principle, suppress the mean-field chiral dynamo phase completely.

In summary, our DNS show that a non-zero constant initial
$\mu$ does not affect chiral dynamos in the non-linear regime, and a very
minor effect is observed if $\mu$ has an initial
sinusoidal spatial variation.
Yet, a systematic exploration of the parameter space and the impact of
initial conditions on a chiral plasma, including the evolution
of $\mu$, is beyond the scope of this paper.

\subsubsection{Effects on collective modes}

\begin{figure}
\begin{center}
  \subfigure{\includegraphics[width=0.32\textwidth]{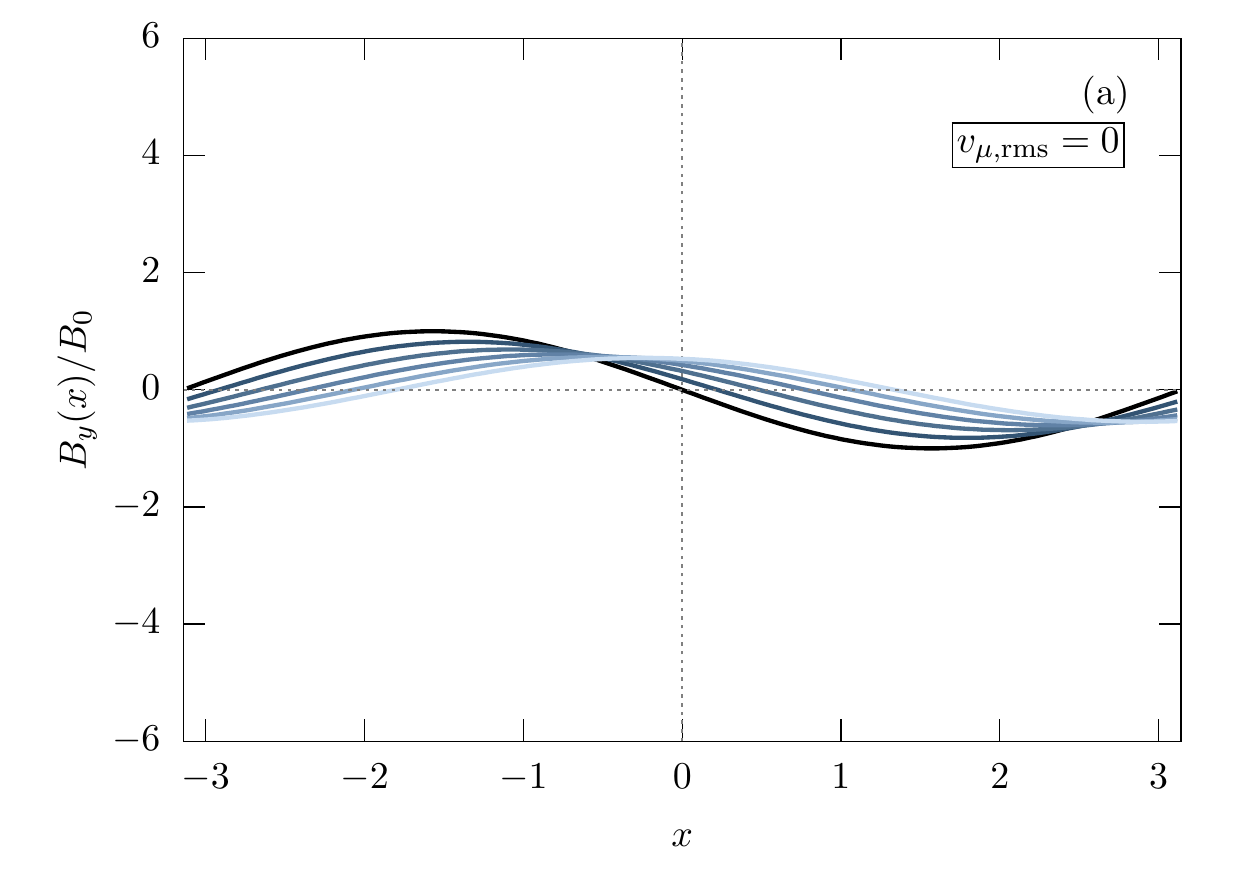}}
  \subfigure{\includegraphics[width=0.32\textwidth]{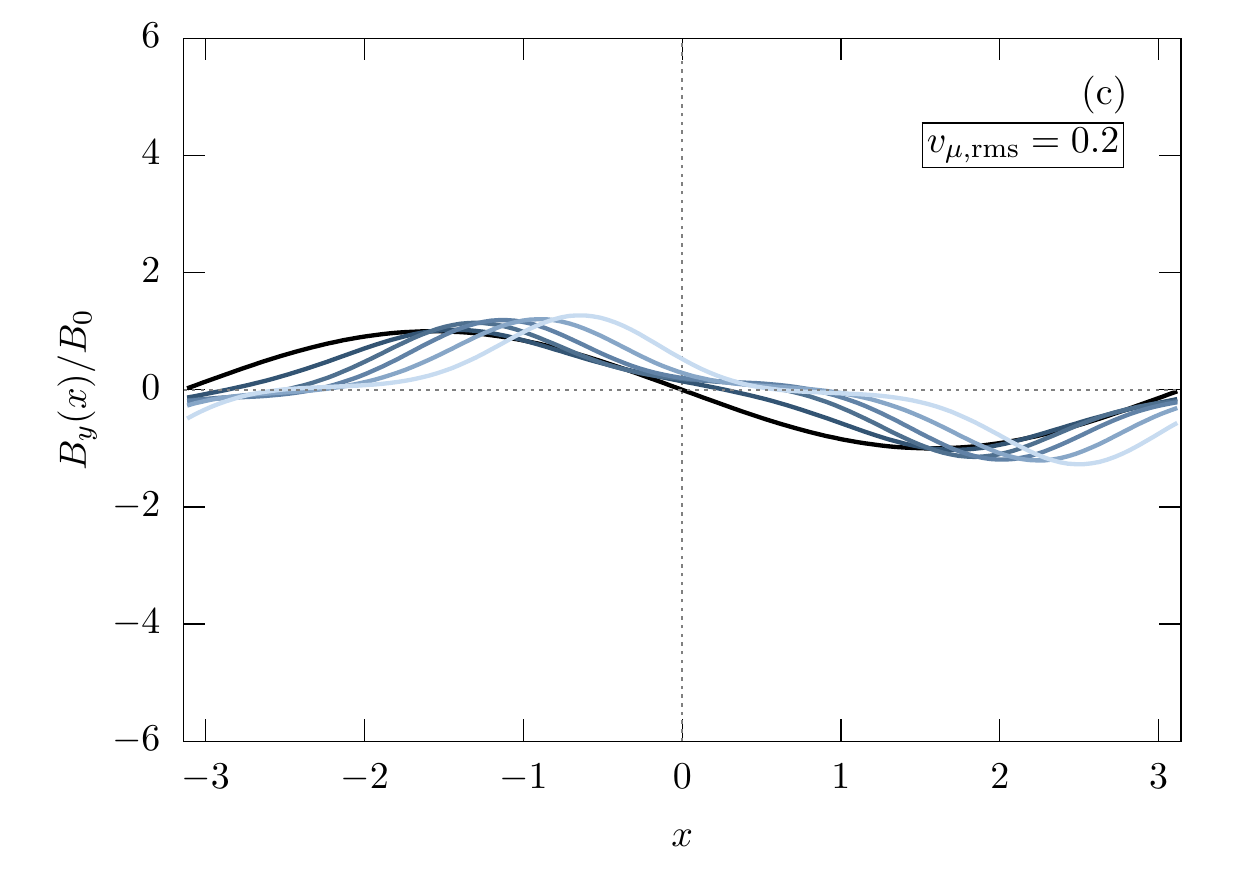}}
  \subfigure{\includegraphics[width=0.32\textwidth]{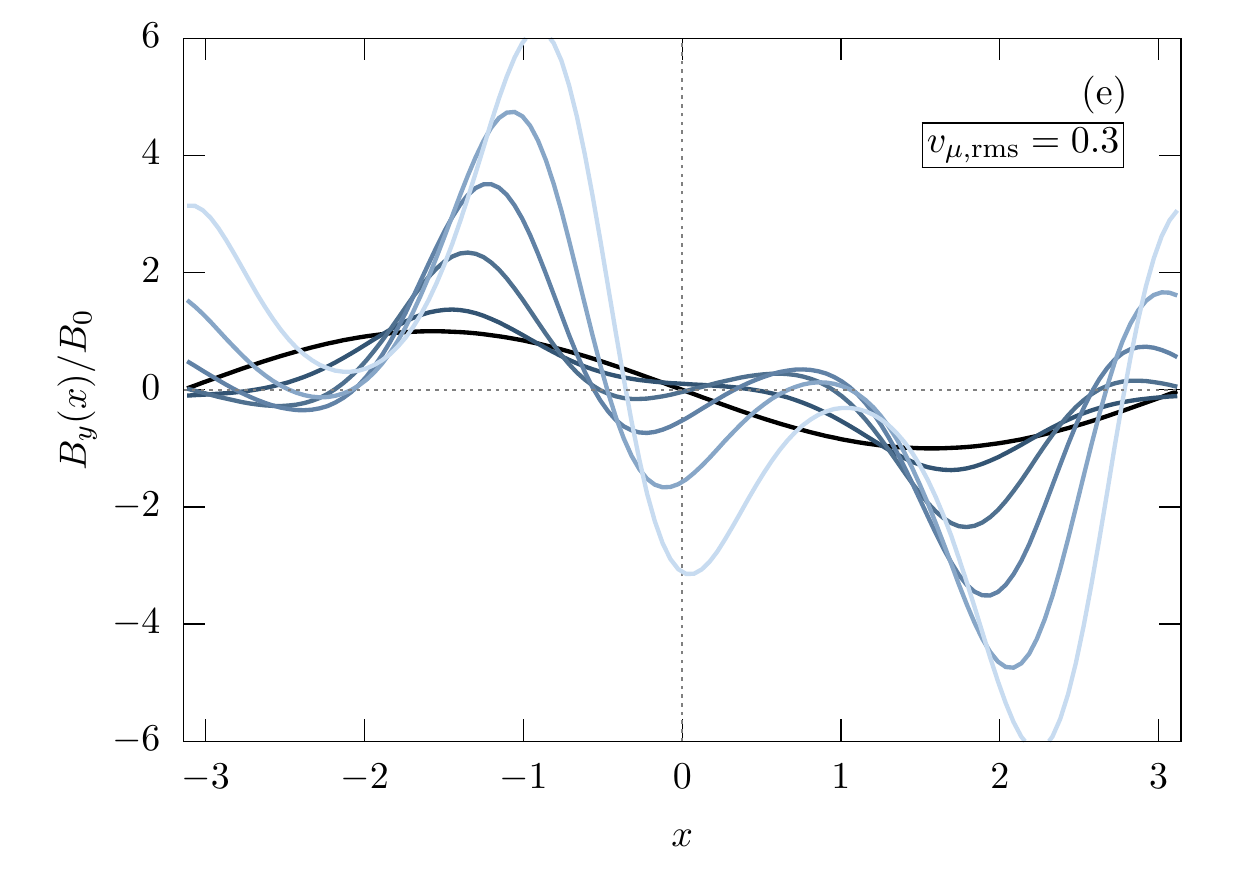}}
  \subfigure{\includegraphics[width=0.32\textwidth]{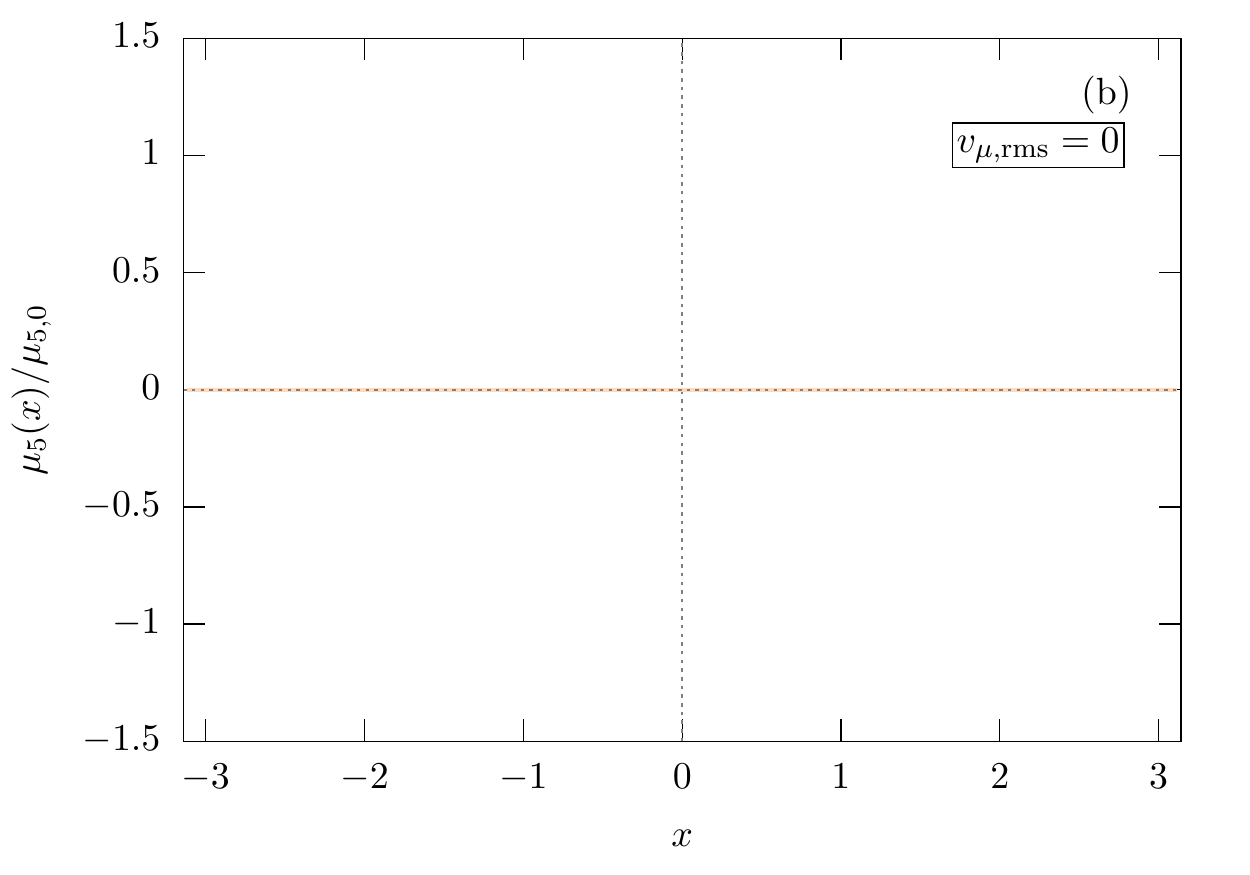}}
  \subfigure{\includegraphics[width=0.32\textwidth]{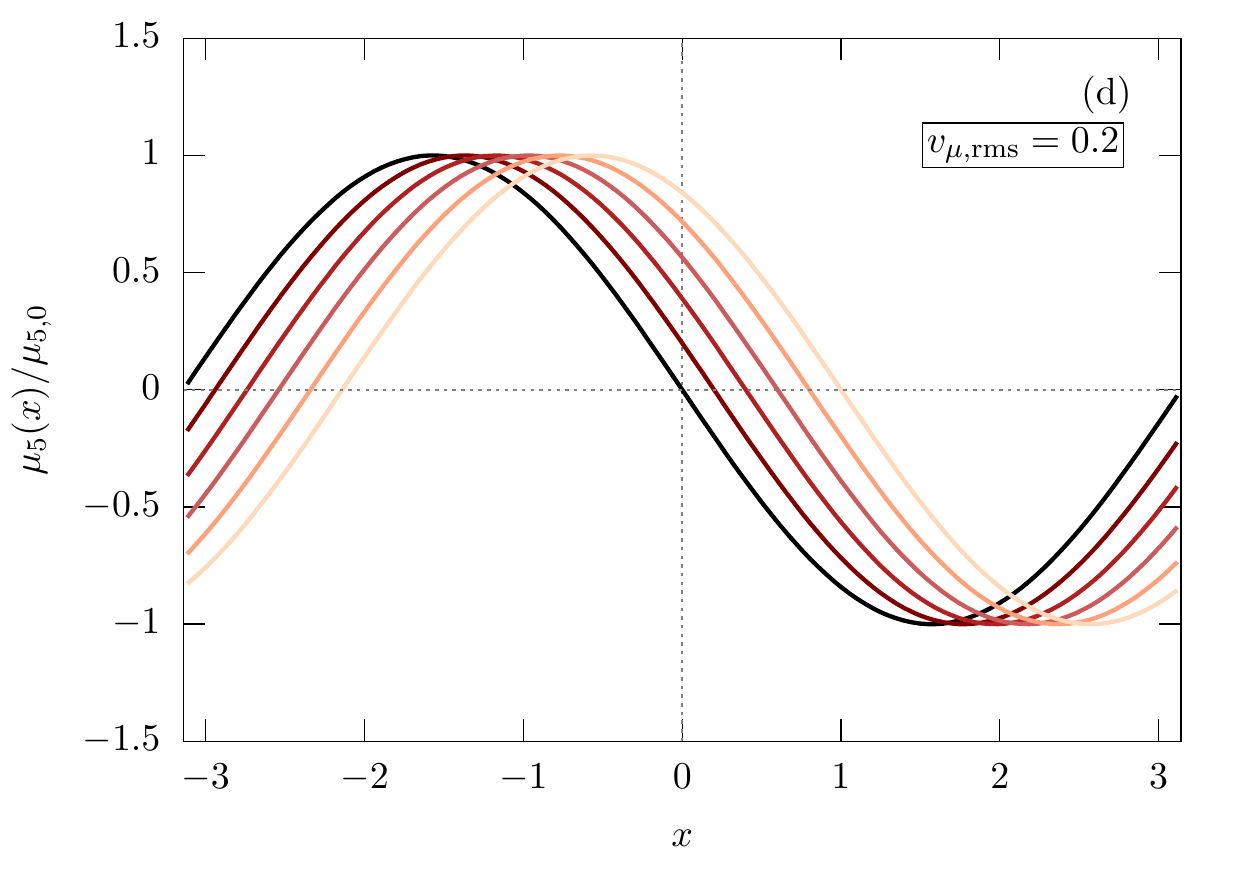}}
  \subfigure{\includegraphics[width=0.32\textwidth]{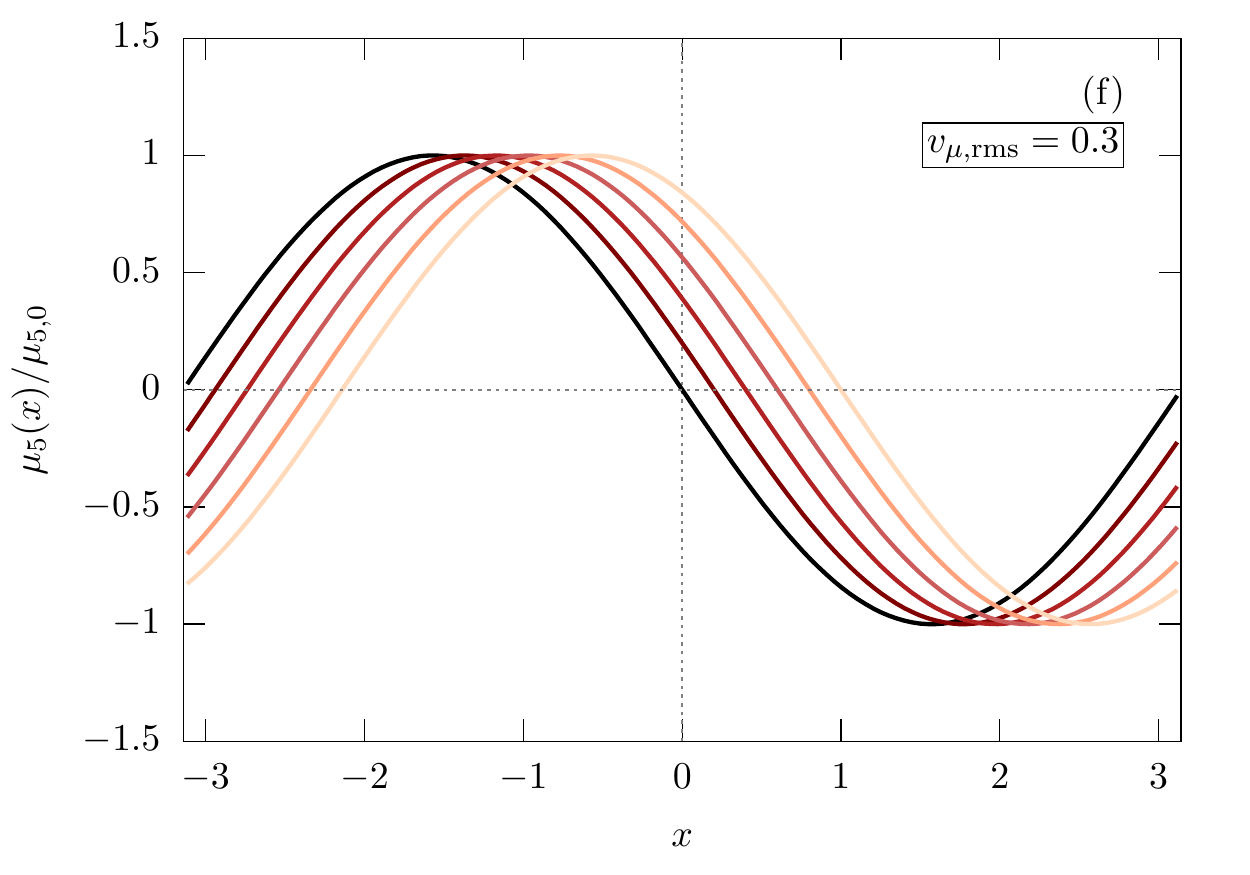}}
\caption{Propagation of a one-dimensional Alfv\'en wave for different
values of $\boldsymbol{v_\mu}$ and with non-zero $\mu$.
The top panels are the same as in figure~\ref{fig_By_x} and in the bottom
panels the propagation of the chiral magnetic wave is shown.
The coupling constants in these runs are $C_5=C_\mu=1$. (Colour online)}
\label{fig_By_x__muS}
\end{center}
\end{figure}

A non-zero chemical potential $\mu$ can trigger chiral magnetic waves \citep[CMWs; see][]{KY11},
as described by the coupled linearised equations~(\ref{mu5-DNS}) and (\ref{mu-DNS}).
The frequency of chiral magnetic waves is
\begin{eqnarray}
   \omega_{\rm CMW} = (C_5 \, C_\mu)^{1/2} \left|{\bm k} {\bm \cdot} {\bm B}_0 \right|.
\label{eq_CMW}
\end{eqnarray}

In order to explore these collective modes, we repeat the 1D runs of
section~\ref{sec_MHDwaves} for a non-zero $\mu_0$.
Again, the initial magnetic field is of the form~(\ref{eq_Alfven1}) and we
use
\begin{eqnarray}
   \mu_{5,0} =  \mu_{5,0}^A \sin(k_x x)
   \quad \mathrm{and} \quad
   \mu_0 = \mu_0^A \sin(k_x x).
\label{eq_muCMW}
\end{eqnarray}
Such an initial condition results in a chiral dynamo
with an $x$-averaged absolute value of the chiral chemical potential
$\mu_{5,\mathrm{rms},0}=2 \mu_{5,0}^A/\pi$, due to the
``quadratic'' nature of the $v_\mu^2$ dynamo.
All the DNS discussed in the following have $\mu_0^A= \mu_{5,0}^A$ and
$C_5=C_\mu$.

In figure~\ref{fig_By_x__muS}, we present results
for $v_{5,\mathrm{rms}}=0$, $v_{5,\mathrm{rms}}=0.2$, and
$v_{5,\mathrm{rms}}=0.3$, using $C_5=C_\mu=1$.
The case of $v_{5,\mathrm{rms}}=0$ (figures~\ref{fig_By_x__muS}(a)--(b)) is
equivalent to figure~\ref{fig_By_x}(a).
For $\mu_{5,0}=0$, the Alfv\'en wave is simply decaying and no CMW
can be observed due to its vanishing amplitude.
A clear difference between figures~\ref{fig_By_x} and \ref{fig_By_x__muS} is
the shape of the magnetic wave.
In the presence of a CMW, the shape of $B_y(x)$ deforms in time, which is most
clearly visible in figure~\ref{fig_By_x__muS}(e).
Here, the chiral dynamo has, on average, the highest growth rate, with
the fastest growth occurring at the location of
the extrema of $\mu_5(x)$.
When the CME and the Alfv\'en wave are out of phase, the $B_y(x)$ curve is
deformed accordingly.

\begin{figure}
\begin{center}
  \includegraphics[width=0.49\textwidth]{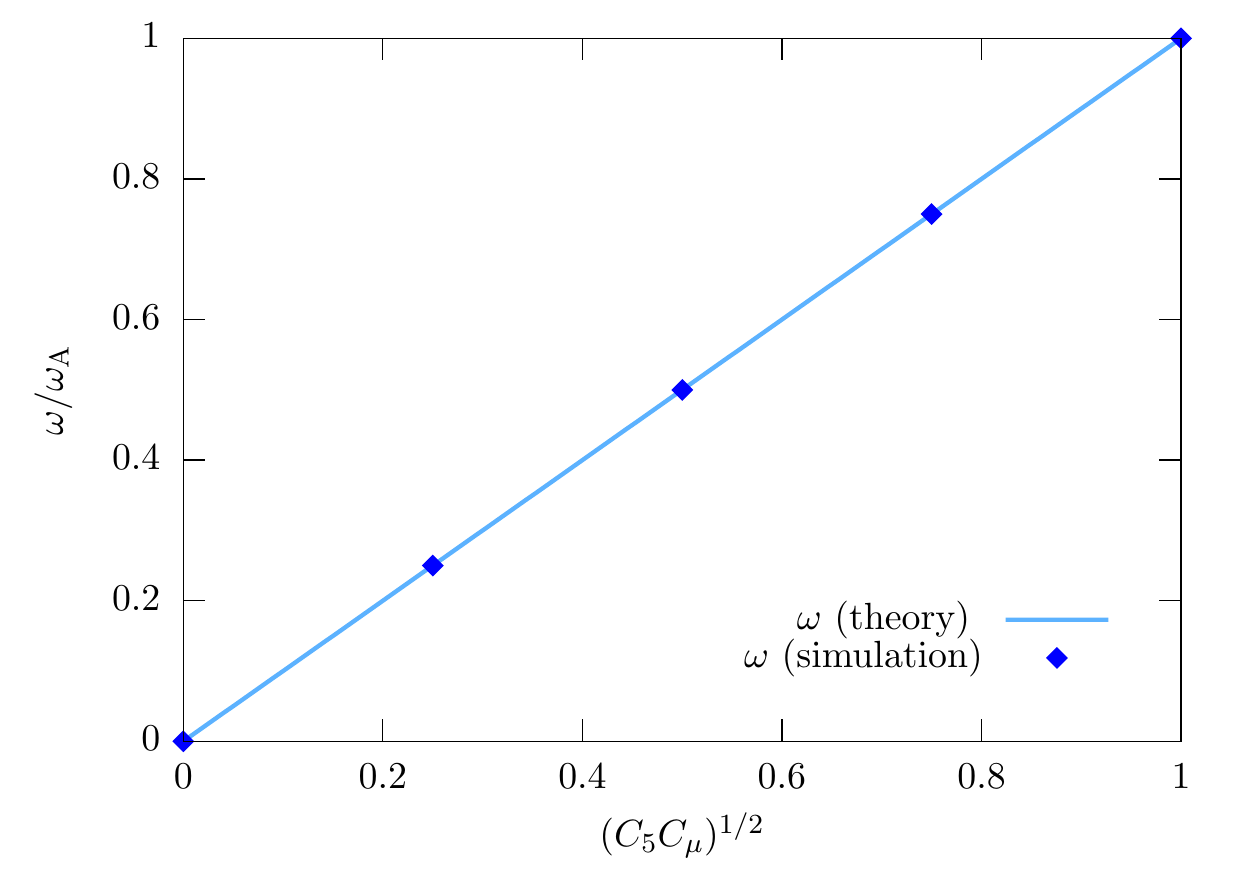}
\caption{Dispersion relation of the chiral magnetic wave for different
values of $(C_5 C_\mu)^{1/2}$.
The simulation setups are same as in the run presented in
figures~\ref{fig_By_x__muS}(c)--(d), in particular $v_\mu=0.2$,
but $(C_5 C_\mu)^{1/2}$ is varied between $0$ and $1$. (Colour online)}
\label{fig_dis_rel__muS}
\end{center}
\end{figure}

For a fixed value of $v_{5,\mathrm{rms}}=0.2$, we change the values
of $C_5$ and $C_\mu$, in order to check the dispersion relation given in
equation~(\ref{eq_CMW}).
As can be seen in figure~\ref{fig_dis_rel__muS}, the 1D simulations agree
perfectly with the theory.
An extended numerical study of CMW and its effects on the nonlinear
evolution of the magnetic field is desirable for future studies.

\section{Conclusions}
\label{sec_conclusions}

Numerical simulations are a key tool for studying the properties of
high-energy plasmas, such as those of the early Universe or of
proto-neutron stars.
At energies $\kB T > 10 \MeV$, the number of degrees of freedom increases by the chiral
chemical potential, which is non-zero in case of an asymmetry between the number
of left- and right-handed fermions.
Through the additional electric current in the presence of such an asymmetry, the
phenomenology of chiral MHD is even richer than that of classical MHD and
numerical simulations are needed to gain a deeper understanding of the
plasma and magnetic field evolution.
To our knowledge, one of the first high-order parallelised codes, which has been
used for chiral MHD, is the \pencilc.
A central purpose of this paper was to describe the implementation of the
chiral MHD module in the \pencilc, to discuss the relevant parameters and
initial conditions in a chiral plasma, and to point out crucial differences to
classical MHD.
We also have presented typical applications of the chiral MHD module
and discussed the obtained numerical results.

First, we have compared the initially laminar dynamo phase
and the dynamo with externally driven turbulence in chiral MHD.
The distinct phases in the two cases were reviewed briefly on the basis of
time series and energy spectra.
We have discussed the mean-field $\alpha_\mu^2$ dynamo,
which can be excited in turbulence via the interaction of magnetic fluctuations
due to tangling of the mean magnetic field by the fluctuating velocity and
magnetic fluctuations produced by the mean chiral chemical
potential.
In DNS, this effect has been seen by measuring the dynamo growth rate
in a stage when turbulence has been produced
by the Lorentz force. Predictions of mean-field
theory for the dynamo growth rate based on the $\alpha_\mu$ effect
are in agreement with the measurements in DNS.

Second, the \pencilc\ was used to check the dispersion relation of
chiral MHD waves and results were compared with analytical predictions.
We find agreement for
the frequencies and the growth or damping rates of the chiral MHD waves:
The chiral dynamo instability leads to a growth of the wave amplitude and
a decrease of the frequency for
chiral velocities larger than the Alfv\'en velocity.

Finally, we have explored the role of the ordinary chemical potential $\mu$
regarding chiral dynamos.
We have demonstrated that $\mu$ can only affect the evolution of $\mu_5$, if
the former has strong gradients.
An initial sinusoidal spatial variation added
to a constant $\mu_0$ can lead to minor
variations of the velocity field in chiral-magnetically driven turbulence.
Additionally, the \pencilc\ was used to study chiral magnetic waves (CMWs), which
occur in the presence of an imposed magnetic field and a non-vanishing coupling
between $\mu$ and $\mu_5$.
As expected, CMWs are decoupled from chiral MHD waves, at least in the
linear regime of the evolution, and their frequency scales with the
square root of the product of the coupling constants, i.e.\ $(C_5 C_\mu)^{1/2}$.

\section*{Acknowledgements}
We are grateful to Dmitri Kharzeev for numerous discussions on
the effects of the chemical potential and chiral magnetic waves
in chiral MHD.
Further, we acknowledge the discussions with participants of the Nordita Scientific Program on
{\it Quantum Anomalies and Chiral Magnetic Phenomena}, Stockholm (September -- October 2018).
The detailed comments on our
manuscript by Matthias Rheinhardt and the anonymous referees
are very much appreciated.
This project has received funding from the
European Union's Horizon 2020 research and
innovation program under the Marie Sk{\l}odowska-Curie grant
No.\ 665667 (``EPFL Fellows'').
We thank for support by the \'Ecole polytechnique f\'ed\'erale de Lausanne, Nordita,
and the University of Colorado through
the George Ellery Hale visiting faculty appointment.
Support through the National Science Foundation Astrophysics and Astronomy Grant Program (grant 1615100),
the Research Council of Norway (FRINATEK grant 231444),
and the European Research Council (grant number 694896) are
gratefully acknowledged.
I.R.\ acknowledges the hospitality of NORDITA,
the Kavli Institute for Theoretical Physics in Santa Barbara
and the \'Ecole Polytechnique F\'ed\'erale de Lausanne.
Simulations presented in this work have been performed
with computing resources
provided by the Swedish National Allocations Committee at the Center for
Parallel Computers at the Royal Institute of Technology in Stockholm.

\newpage

\bibliographystyle{gGAF.bst}

\markboth{J.~SCHOBER ET AL.}{GEOPHYSICAL AND ASTROPHYSICAL FLUID DYNAMICS}
\appendix
\markboth{J.~SCHOBER ET AL.}{GEOPHYSICAL AND ASTROPHYSICAL FLUID DYNAMICS}

\section{Chiral MHD equations in dimensionless form}
\label{appendix_dimform}

For DNS, it is convenient to move from a system formulated in physical units to
a dimensionless one.
This can be achieved when velocity is measured in units of the sound speed
$c_\mathrm{s}$, length is measured in units of $\mu_{5,0}^{-1}$,
where $\mu_{5,0}$ is the initial value of a uniform $\mu_5$, and
time is measured in units of $(c_\mathrm{s}\mu_{5,0})^{-1}$.
With the definitions ${\bm \BB}=\sqrt{\meanrho} \, c_\mathrm{s}\tilde{\bm \BB}$,
${\bm \UU}=c_\mathrm{s}\tilde{\bm \UU}$, $\mu_5=\mu_{5,0} \,\tilde{\mu}_5$,
$\mu=\mu_{5,0} \, \tilde{\mu}$,
and $\rho= \meanrho \, \tilde\rho$, where $\meanrho$ is the volume-averaged
density, the system of
equations~(\ref{ind-DNS})--(\ref{mu-DNS}) can be written as
\begin{eqnarray}
\frac{\upartial \tilde\BB}{\upartial \tilde t}  &=&   \tilde{\bm \nabla}
   {\bm \times} \biggl[\tilde{\bm \UU} {\bm \times} \tilde{\BB}  + {\rm Ma}_\mu \,
   \Big(\tilde{\mu}_5 \tilde{\BB}  - \tilde{\bm \nabla} {\bm \times} {\tilde\BB} \Big)\biggr],
\label{ind-NS}\\
\tilde \rho {{\mathrm D} \tilde \UU \over {\mathrm D} \tilde t}&=& (\tilde \nab \times \tilde{\BB})  \times
   \tilde \BB -\tilde \nab \tilde p + {\rm Re}_5^{-1} \tilde\nab {\bm \cdot}
   (2 \tilde \rho \tilde \SSSS)
   + \tilde \rho \tilde\ff ,
\label{UU-NS}\\
\frac{{\mathrm D} \tilde\rho}{{\mathrm  D} \tilde t} &=& - \tilde\rho \, \tilde\nab  \cdot \tilde\UU ,
\label{rho-NS}\\
\frac{{\mathrm D} \tilde{\mu}_5}{{\mathrm D} \tilde t}  &=& \tilde D_5 \, \tilde\Delta \tilde{\mu}_5
   + \Lambda_5 \, \Big[{\tilde\BB} {\bm \cdot} (\tilde\nab {\bm \times}
   {\tilde\BB}) - \tilde{\mu}_5 {\tilde\BB}^2 \Big] - \tilde \Gamma_\mathrm{f} \tilde{\mu}_5
   -\tilde C_5 (\tilde\BB {\bm \cdot} \tilde\nab) \tilde \mu,
\label{mu5-NS} \\
\frac{{\mathrm D} \tilde{\mu}}{{\mathrm D} \tilde t}  &=& \tilde D_\mu \, \tilde\Delta \tilde{\mu}
   -\tilde C_\mu (\tilde\BB {\bm \cdot} \tilde\nab) \tilde \mu_5
\label{mu-NS}.
\end{eqnarray}
A summary of the chiral parameters and their names in the \pencilc\ can be
found in Table~\ref{tab_chiralMHDparameters}.
We have introduced the following dimensionless parameters.
\begin{itemize}
\item{The \textit{chiral Mach number}
\begin{eqnarray}
  {\rm Ma}_\mu = {\eta\mu_{5,0}}\big/{c_\mathrm{s}} \equiv {v_\mu}\big/{c_\mathrm{s}} ,
\label{Ma_mu_def}
\end{eqnarray}
which measures the relevance
of the chiral term in the induction
equation~(\ref{ind-DNS}) and determines
the growth rate of the small-scale chiral dynamo instability.
}
\item{The \textit{magnetic Prandtl number}
\begin{eqnarray}
  \PrM = {\nu}/{\eta} ,
\end{eqnarray}
which is equivalent to the definition in classical MHD.
}
\item{The \textit{chiral Prandtl number}
\begin{eqnarray}
  {\rm Pr}_{_{5}} = {\nu}\big/{D_5},
\end{eqnarray}
which measures the ratio of viscosity and diffusion of $\mu_5$.
}
\item{The \textit{chemical potential Prandtl number}
\begin{eqnarray}
  {\rm Pr}_\mu = {\nu}\big/{D_\mu},
\end{eqnarray}
which measures the ratio of viscosity and diffusion of $\mu$.
}
\item{The \textit{chiral nonlinearity parameter}
\begin{eqnarray}
  \lambda_5 = \lambda \eta^2 \meanrho ,
\label{eq_lambamu}
\end{eqnarray}
which characterises the nonlinear back reaction
of the magnetic field on the chiral chemical potential $\tilde{\mu}_5$.
The value of $\lambda_5$
affects the strength of the saturation magnetic field and
the strength of the magnetically driven turbulence.
}
\item{The \textit{chiral flipping parameter}
\begin{eqnarray}
  \tilde\Gamma_\mathrm{f} = {\Gamma_\mathrm{f}}\big/{\big(\mu_{5,0} c_\mathrm{s}\bigr)} ,
\end{eqnarray}
which measures the relative importance
of chiral flipping reactions.
}
\item{The \textit{coupling parameters}
\begin{eqnarray}
  \tilde C_5 =  \sqrt{\meanrho}  \, C_5
\end{eqnarray}
and
\begin{eqnarray}
  \tilde C_\mu =  \sqrt{\meanrho} \,  C_\mu ,
\end{eqnarray}
which measure the strength of the coupling between the evolution of
$\mu$ and $\mu_5$, respectively.
}
\end{itemize}
Using the definitions above, one finds that $\tilde D_5={\rm Ma}_\mu \, \PrM
/{\rm Pr}_{_{5}}$,
$\tilde D_\mu={\rm Ma}_\mu \, \PrM / {\rm Pr}_\mu$,
$\, \Lambda_5 =\lambda_5/{\rm Ma}_\mu$,
and ${\rm Re}_5 = \left({\rm Ma}_\mu \, \PrM\right)^{-1}$ in
equations~(\ref{ind-NS})--(\ref{mu-NS}).
\begin{table}
  \centering
  \caption{Chiral MHD parameters in the \pencilc}
  \label{tab_chiralMHDparameters}
  \begin{tabular}{lll}
  \hline
  \hline
    Dimensionless parameter 		& Name in the \pencilc     \\
  \hline
  	$\tilde{\mu}_5$      		& \texttt{p\%mu5} \\
  	$\tilde{\mu}_S$      		& \texttt{p\%muS} \\
	$\Lambda_5$			& \texttt{lambda5} \\
        $\tilde D_5$			& \texttt{diffmu5} \\
        $\tilde D_\mu$			& \texttt{diffmuS} \\
        $\tilde C_5$			& \texttt{coef\_mu5} \\
        $\tilde C_\mu$			& \texttt{coef\_muS} \\
	$\tilde\Gamma_\mathrm{f}$	& \texttt{gammaf5} \\
  \hline
  \hline
  \end{tabular}
\end{table}

\section{A chiral MHD setup in the {\bfseries{\scshape Pencil Code}} }
\label{appendix_Makefile}

An example for a minimum set up of the \texttt{src/Makefile.local} looks like
this:
\begin{verbatim}
###                             -*-Makefile-*-
### Makefile for modular pencil code -- local part
### Included by `Makefile'
###

MPICOMM        =  nompicomm
HYDRO          =    hydro
DENSITY        =    density
MAGNETIC       =    magnetic
FORCING        =  noforcing
VISCOSITY      =    viscosity
EOS            =    eos_idealgas
SPECIAL        =    special/chiral_mhd
REAL_PRECISION =    double
\end{verbatim}

Further, for running the chiral MHD module, one needs to add
\begin{verbatim}
   &special_init_pars
   initspecial='const', mu5_const=10.
\end{verbatim}
to \texttt{start.in} and
\begin{verbatim}
   &special_run_pars
   diffmu5=1e-4, lambda5=1e3, cdtchiral=1.0
\end{verbatim}
to \texttt{run.in}, where we have chosen exemplary values for the chiral
parameters.

\end{document}